%% file: main.tex
\documentclass[Afour,sageh,times]{sagej}
\usepackage{luatex85}

\usepackage{moreverb,url}

\usepackage[colorlinks,bookmarksopen,bookmarksnumbered,citecolor=red,urlcolor=red]{hyperref}
\usepackage{multirow}

\newcommand\BibTeX{{\rmfamily B\kern-.05em \textsc{i\kern-.025em b}\kern-.08em
T\kern-.1667em\lower.7ex\hbox{E}\kern-.125emX}}

\def\volumeyear{2025}

\usepackage{balance}
\usepackage[T1]{fontenc}
\usepackage[utf8]{inputenc}
\usepackage{microtype}
\usepackage{graphicx}
\usepackage[svgnames,rgb,x11names]{xcolor}

\usepackage{booktabs} 
\usepackage{algorithm}
\usepackage{algpseudocode}
\algrenewcommand{\algorithmiccomment}[1]{\hfill\emph{> #1}}
\usepackage{minted}
\newlength{\mintednumbersep}
\AtBeginDocument{%
  \sbox0{\tiny00}%
  \setlength\mintednumbersep{\leftskip}%
  \addtolength\mintednumbersep{-\wd0}%
}

\usepackage{tikz}
\usetikzlibrary{angles,quotes}
\usetikzlibrary{tikzmark}
\usetikzlibrary{arrows.meta}
\usepackage{siunitx}
\usepackage{todonotes}

\usepackage[a-1b]{pdfx}

\definecolor{dkgreen}{RGB}{0,64,0}
\definecolor{ltgray}{RGB}{245,245,245}
\definecolor{mauve}{RGB}{139,0,139}

\usepackage{caption}
\usepackage{subcaption}
\usepackage{enumerate}

\usepackage{multirow}
\usepackage{array}
\usepackage{mwe}
\usepackage{keycommand}
\usepackage{fontawesome5}
\usepackage{amssymb}
\selectcolormodel{natural} %
\usepackage{ninecolors} %
\selectcolormodel{rgb} %
\usepackage{tabularray}
\UseTblrLibrary{booktabs}
\usepackage{makecell}
\usepackage{enumitem}
\usepackage{orcidlink}
\usepackage{comment}
\usepackage{tcolorbox}

\makeatletter
\newcommand{\thickhline}{%
    \noalign {\ifnum 0=`}\fi \hrule height 1pt
    \futurelet \reserved@a \@xhline
}
\newcolumntype{"}{@{\hskip\tabcolsep\vrule width 1pt\hskip\tabcolsep}}
\makeatother

\newtcolorbox{boxE}{
    colback = white,
    colframe = lightgray,
    sharp corners=all
}

\newcommand{\rev}[1]{\textcolor{MediumPurple2}{#1}}
\renewcommand{\rev}[1]{#1}
\newcommand{\supp}[1]{\textcolor{MediumSeaGreen}{#1}}
\renewcommand{\supp}[1]{#1}

\newkeycommand\benchtagBackend[bg=lightgray, fg=white, icon=\faIcon{tag}][1]{\tikz[baseline=(X.base), inner sep=1pt, outer sep=1pt, text depth=0.25ex, text height=1.3ex]{\node (X) [font=\scriptsize, fill=\commandkey{bg}, text=\commandkey{fg}] {\raisebox{0.1ex}{\resizebox{4pt}{!}{\commandkey{icon}}}\,#1};}}

\newcommand\rrrrightarrows{%
  \mathrel{{\ooalign{\hss\raisebox{-0.2ex}{$\rightrightarrows$}\hss\cr\raisebox{1.3ex}{$\rightrightarrows$}}}}
}

\colorlet{appDomain}{LightSkyBlue}
\colorlet{scale}{LightGreen!70!SlateGrey}
\colorlet{communication}{LightCoral}
\colorlet{programmingmodel}{LightSalmon}
\colorlet{programminglanguage}{LightGoldenrod}
\colorlet{memaccess}{MediumPurple!70!white}
\colorlet{method}{LightSeaGreen}
\colorlet{computeperf}{DarkSeaGreen}
\colorlet{communicationperf}{communication!60!SlateGrey}
\colorlet{mesh}{memaccess!60!SlateGrey}
\newcommand\btAppDomain[1]{\benchtagBackend[bg=appDomain,icon=\faIcon{flask}]{#1}}
\newcommand\btScale[1]{\benchtagBackend[bg=scale,icon=\faIcon{chart-line}]{#1}}
\newcommand\btComm[1]{\benchtagBackend[bg=communication,icon=\faIcon{network-wired}]{#1}}
\newcommand\btProgMod[1]{\benchtagBackend[bg=programmingmodel,icon=$\rrrrightarrows$]{#1}}
\newcommand\btProgLang[1]{\benchtagBackend[bg=programminglanguage,icon=\faIcon{code}]{#1}}
\newcommand\btMemAcc[1]{\benchtagBackend[bg=memaccess,icon=\faIcon{memory}]{#1}}
\newcommand\btMethod[1]{\benchtagBackend[bg=method,icon=\faIcon{cog}]{#1}}
\newcommand\btCompPerf[1]{\benchtagBackend[bg=computeperf,icon=\faIcon{calculator}]{#1}}
\newcommand\btCommPerf[1]{\benchtagBackend[bg=communicationperf,icon=\faIcon{comments}]{#1}}
\newcommand\btMesh[1]{\benchtagBackend[bg=mesh,icon=\faIcon{th}]{#1}}
\newcommand\btMath[1]{\benchtagBackend[bg=mesh,icon=\faIcon{university}]{#1}}

\newcommand{\licBsdThree}{BSD-3-Clause}

\tikzset{
  license/.style={circle, draw, minimum width=3em, font=\sffamily\bfseries, very thick},
  license version/.style={circle, fill=black, text=white, font=\sffamily\bfseries\footnotesize, inner sep=1.5pt}
}%
\newcommand{\anyLicense}[1]{%
\scalebox{0.5}{%
\tikz[inner sep=0pt, outer sep=0pt, baseline=(lic.base)]{\node (lic) [license] {#1};}%
}}%
\newcommand{\anyLicenseVersion}[2]{%
\scalebox{0.5}{%
\tikz[inner sep=0pt, outer sep=0pt, baseline=(lic.base)]{\node (lic) [license] {#1};\node [shift=(lic.north east), license version] {#2};}%
}}%

\newcommand{\licMit}{\anyLicense{MIT}}
\newcommand{\licNone}{\anyLicense{\emph{None}}}
\newcommand{\licFree}{\anyLicense{\emph{Free}}}
\newcommand{\licProp}{\anyLicense{\emph{Prop}}}
\newcommand{\licGpl}{\anyLicense{GPL}}
\newcommand{\licGplThree}{\anyLicenseVersion{GPL}{3}}
\newcommand{\licGplTwo}{\anyLicenseVersion{GPL}{2}}
\newcommand{\licLgplTwoone}{\anyLicenseVersion{LGPL}{2}}
\newcommand{\licApacheTwo}{\anyLicenseVersion{Apch}{2}}
\newcommand{\licMplTwo}{\anyLicenseVersion{MPL}{2}}
\newcommand{\licCcByFour}{\anyLicenseVersion{CC-B}{4}}
\renewcommand{\licBsdThree}{%
\scalebox{0.5}{%
\tikz[inner sep=0pt, outer sep=0pt, baseline=(lic.base)]{\node (lic) [license] {BSD};\node [shift=(lic.south east), license version] {3};}%
}}%
\newcommand{\licBsdFour}{%
\scalebox{0.5}{%
\tikz[inner sep=0pt, outer sep=0pt, baseline=(lic.base)]{\node (lic) [license] {BSD};\node [shift=(lic.south east), license version] {4};}%
}}%
\newcommand{\licBsdTwo}{%
\scalebox{0.5}{%
\tikz[inner sep=0pt, outer sep=0pt, baseline=(lic.base)]{\node (lic) [license] {BSD};\node [shift=(lic.south east), license version] {2};}%
}}%

\title{%
    An HPC Benchmark Survey and Taxonomy for Characterization%
}

\date{December 2024}

\begin{document}
\runninghead{Herten, Pearce, and Guimar\~aes}

\author{Andreas Herten\orcidlink{0000-0002-7150-2505}\affilnum{1}, 
Olga Pearce\orcidlink{0000-0002-1904-9627}\affilnum{2,3}, Filipe S. M. Guimar\~aes\orcidlink{0000-0002-5618-6727}\affilnum{1}}

\affiliation{\affilnum{1}Forschungszentrum J\"ulich, J\"ulich, Germany, a.herten@fz-juelich.de 
\\\affilnum{2}Lawrence Livermore National Laboratory, Livermore, CA, USA
\\\affilnum{3}Texas A\&M University, College Station, TX, USA}

\corrauth{Olga Pearce}
\email{olga@llnl.gov}

\input{sections/1-abstract}


\maketitle

\input{sections/2-intro}

\input{sections/3-related}

\input{sections/4-taxonomy}

\input{sections/5-benchmarks}

\input{sections/6-evaluation}

\input{sections/7-conclusions}
\input{sections/8-ack}
\input{sections/9-bio}

\bibliographystyle{SageH}
\bibliography{cite,benchmarkrefs,benchmarklinks}

\onecolumn
\input{sections/benchmark_table-smallC}
{
\ExplSyntaxOn
\NewDocumentCommand\FirstOfTwo{m}{
  \exp_last_unbraced:Ne \use_i:nn {#1}
}
\NewDocumentCommand\SecondOfTwo{m}{
  \exp_last_unbraced:Ne \use_ii:nn {#1}
}
\ExplSyntaxOff
\DefTblrTemplate{caption-text}{first}{\FirstOfTwo{\InsertTblrText{caption}}}
\DefTblrTemplate{caption-text}{second}{\SecondOfTwo{\InsertTblrText{caption}}}
\DefTblrTemplate{firsthead}{default}{
  \centering
  \UseTblrTemplate{caption-tag}{default}%
  \UseTblrTemplate{caption-sep}{default}%
  \UseTblrTemplate{caption-text}{first}
  \par
}
\DefTblrTemplate{middlehead,lasthead}{default}{
  \centering
  \UseTblrTemplate{caption-tag}{default}%
  \UseTblrTemplate{caption-sep}{default}%
  \UseTblrTemplate{caption-text}{second}
  \par
}
\appendix
\setcounter{table}{0}
\renewcommand{\thetable}{A\arabic{table}}
\input{sections/benchmark_table-full}
}
\twocolumn

\end{document}

%% file: sections/1-abstract.tex
\begin{abstract}
The field of High-Performance Computing (HPC) is defined by providing computing devices with highest performance for a variety of demanding scientific users. The tight co-design relationship between HPC providers and users propels the field forward, paired with technological improvements, achieving continuously higher performance and resource utilization. A key device for system architects, architecture researchers, and scientific users are benchmarks, allowing for well-defined assessment of hardware, software, and algorithms. Many benchmarks exist in the community, from individual niche benchmarks testing specific features, to large-scale benchmark suites for whole procurements. We survey the available HPC benchmarks, summarizing them in table form with key details and concise categorization\supp{, also through an interactive website}. For categorization, we present a \emph{benchmark taxonomy} for well-defined characterization of benchmarks.
\end{abstract}

%% file: sections/2-intro.tex
\section{Introduction}
\label{sec:intro}

High-Performance Computing (HPC) is, by definition, aiming for excellent performance of the executed workloads. State-of-the-art hardware is deployed for scientists, engineers, and other researchers with ever-increasing performance capabilities. At the same time, these users solve computational challenges of ever-increasing sophistication, driving the demand for HPC systems.

While some expert users might understand the performance characteristics and hardware-software interplay of their applications, that is not generally the case for most users, system administration, HPC engineers, and support staff. Hence, dedicated applications with well-defined workloads are utilized to assess the performance of systems. These benchmarks have a long tradition in the HPC field, with the High-Performance LINPACK (HPL, ~\cite{hpl_ref}), used to rank supercomputers in the Top500\footnote{\url{https://www.top500.org}}, arguably the most prominent.
But many benchmarks exist, focusing on diverse aspects of hardware and software of HPC installations: Individual niche benchmarks test dedicated hardware characteristics, other more integrated programs test interplay of different components, and some benchmarks entirely focus on software components. The choice is plentiful, but keeping track is hard.

\subsection{Profile of Benchmarks}

For good reason, benchmarks are one of the key elements in the HPC researcher/engineer toolbox. Their benefit has many layers:

\begin{description}
    \item[\rev{Clarity}] Through the well-defined setup, workloads, and execution instruction, benchmarks allow objective, repeatable, and transparent performance measurements; the key is a clear -- and ideally simple -- metric.
    \item[Comparability] Analysis of various hardware installations with identical or similar benchmark execution makes the installations comparable and assessable by means of the benchmark's metric.
    \item[\rev{Durability}] \rev{Through well-designed benchmarks, an accessible and continuous assessment of systems is enabled, allowing for tracking historical data and understanding technological improvements and trends.}
    \item[Advancement] \rev{Modeling the complex interplay of hardware and software, benchmarks enable} focused hardware research and application development \rev{towards ecosystem advancement}; especially in relation to the theoretical capabilities of hardware (\emph{peak performance}). \rev{A} benchmark results database can create a competitive drive for system improvement.
    \item[\rev{Decisiveness}] Ultimately, analyzing robust benchmarks across different hardware installations over time allows informed, objective decisions about system investment.  Benchmarks are key in modern system procurements, where execution performance counts more than the \rev{hardware peak performance}.
    \item[Validation] Well-defined benchmarks allow tracking resource utilization and performance regressions on in-production systems; for new systems, well-known benchmarks can be used to validate the system for production; they can also be used for stress-testing systems and support reliability of the system.
\end{description}

The benefit of individual benchmarks is amplified when they are collected into benchmark suites. Various workloads are combined to conduct holistic system characterization, usually with normalized, comparable metrics.

Benchmarks not only have value for HPC researchers and engineers, but also for HPC users. By having a well-defined version of their HPC application, users may track performance regressions in their program and understand performance limiters. By supplying benchmarks to the HPC community, users have a direct impact on system design and procurement decisions. In turn, seeing benchmark results for HPC systems, users have comparable baselines and can gain confidence in the capabilities of the system.

Of course, benchmarks have many limitations. Comparability crossing hardware generations and vendors is a hard task, as highly-optimized HPC applications optimize for specific hardware, which is a priori not directly transferable. A certain benchmark metric may be valuable, but might not convey general information about a system -- especially synthetic benchmarks tend to measure specific aspects of a system, which have limited real-world applicability for more involved applications. Finally, creating a robust, repeatable, portable, versatile, stable, and clear benchmark is a very involved task, which requires significant effort.

\subsection{Contribution}

Because of their importance, the HPC field created a vast set of different benchmarks over the years. They range from simple synthetic benchmarks, over mini-apps of scientific applications, to full-blown large-scale applications with possibly large input files. Some benchmarks are collected into benchmark suites, typically created for system procurements, to replicate a desired measurement and workload mix.

Some benchmarks are well-known in the field -- like HPL or STREAM~\citep{stream_ref} -- while others are only known to a small subset. The benchmarks/suites may be published only on websites accompanying a procurement, or are hosted in a GitHub repository attached to a journal publication; finding them can be challenging.

To improve their visibility, findability, and, ultimately, benefit for the field, we contribute in this work a \textbf{collection of benchmarks and benchmark suites}. In the survey, we present each benchmark and suite with metadata, including license, URL for download, reference, notes, and categorization. The survey consists of \supp{a summarizing overview table}, added at the end of this article\supp{, a supplemental table with all details and categories available alongside the article online}, a dynamic \supp{website available online}, and an evaluation of the collected material. For categorization, we develop a \textbf{Benchmark Taxonomy}, allowing for well-defined characterization of each benchmark and quick assessment of suitability. While taking great care to capture most of the benchmarks and suites of the field, it is likely that individual, potentially more niche benchmarks are missing. To that extent, we publish the benchmark survey in raw form\footnote{The data is available in simple, machine-readable YAML form, for which we developed a fitting schema. The YAML data is the single source of truth of \rev{which} the PDF and HTML tables are generated from; all required scripts are also part of the repository.} and the taxonomy keywords on GitHub as open source software/data for future extension. The \supp{interactive} online version of the table is available at \href{https://fzj-jsc.github.io/benchmark-survey/}{\url{fzj-jsc.github.io/benchmark-survey/}}, the raw benchmark list is available at \href{https://github.com/FZJ-JSC/benchmark-survey/}{\url{github.com/FZJ-JSC/benchmark-survey/}}, and the taxonomy definition at \href{https://github.com/LLNL/benchpark/blob/develop/taxonomy.yaml}{\url{github.com/LLNL/benchpark/blob/develop/taxonomy.yaml}}.

\subsection{Structure}

The rest of the paper is structured as follows. In \autoref{sec:related} we concisely assess the state-of-the-field and discuss related work. In \autoref{sec:taxonomy}, we present the Benchmark Taxonomy. Finally, in \autoref{sec:benchmarks} we present the benchmarks in table form. Some observations and evaluations are discussed in \autoref{sec:eval}. In \autoref{sec:conclusions}, we conclude our paper.

%% file: sections/3-related.tex
\section{Related Work}
\label{sec:related}

Efforts to publicly post performance-focused benchmark results relating to deployed HPC systems 
range from the long-running Top500 and aligned Green500\footnote{\url{https://top500.org/lists/green500}} and IO500\footnote{\url{https://io500.org/}}, 
to sub-discipline benchmarking such as Machine Learning Commons (MLCommons, ~\cite{mlperf,mattson2019mlperf}) and HPL-MxP\footnote{\url{https://hpl-mxp.org}}. Notable is also \href{https://openbenchmarking.org/}{OpenBenchmarking.org}, which openly collects results from the Phoronix Test Suite\footnote{\url{https://www.phoronix-test-suite.com/}} and schema-compliant further benchmarks; the focus is on end-user devices.

There are numerous efforts to make benchmarking easier,
by encoding build/run/evaluation rules for entire suites of benchmarks;
examples include Pavilion~\citep{pavilion}, Reframe~\citep{10.1007/978-3-030-44728-1_3},
JUBE~\citep{Breuer2022JUBE}, Ramble~\citep{ramble-hpctests2023},
and Benchpark~\citep{hpctests2023-benchpark}.
While they showcase somewhat different approaches,
the proliferation of such efforts underscores
the vital importance -- and difficulty -- of benchmarking.

There have of course been attempts at surveying the HPC benchmarks\rev{ -- \emph{meta-benchmarking}, in a sense}. Here we list just a few of the recent ones.
A survey of convergence of big data, HPC, and ML 
systems has been done by \cite{10.1007/978-3-030-94437-7_7} and
cites 25 benchmarks/suites (our paper covers the HPC-specific
ones from this list).
In similar spirit list \cite{Thiyagalingam_2022} a number of scientific ML benchmarks and present the SciMLBench framework.
Some works argue that although we already have too many benchmarks
in the ML space~\citep{10.1007/978-3-030-32813-9_5} (citing 54 
ML benchmarking papers), we still need more - while needing convergence.
This underscores the need for the community awareness of 
existing benchmarking work, better benchmark characterization,
and, hopefully, collaboration to both improve the quality of benchmarks,
and understand their applicability.

%% file: sections/4-taxonomy.tex
\section{Taxonomy}
\label{sec:taxonomy}
Benchmarks come in different flavors with different execution profiles, focus points, dependencies, and other properties. While comparing well-known benchmarks colloquially is easy, a thorough, structured comparison is more involved -- especially for more niche and specialized benchmarks.

In our survey, we identify the different flavors and develop a systematic approach for characterization of benchmarks, considering a variety of different aspects of the individual benchmarks. This \textbf{Benchmark Taxonomy} consists of a number of \emph{categories} under which benchmarks can be sorted and which group different \emph{entries} of the similar kind. A combination of multiple category-entry pairs allows for fine-grained description of each benchmark.

Categories range from the application domain, where a benchmark originates from (like astrophysics), over the method employed for execution (like FFT), and the employed programming language (like Fortran), to detailed aspects like memory access characteristics (like regular access).

The categories are presented in \autoref{fig:tax} with all currently collected entries (normalized to lower-case). The list of possible entries is likely not complete for every possible workload; it rather represents the result of our collection, augmented with other obvious entries.

To allow for extension of the taxonomy and support future work building up on it, the raw data is available in concise YAML form 
\href{github.com/LLNL/benchpark/blob/develop/taxonomy.yaml}{on GitHub}. The schema employed uses the categories for keys, and individual entries available as a list for the values; for example \mintinline{yaml}{communication: [mpi, nccl]}.

Each category has an associated symbol and color, which is used in \emph{tag} form within \supp{the full version of the table (\autoref{table:benchmarks}) and in the interactive online version}. An example tag for a benchmark utilizing MPI communication is \btComm{mpi}. In the machine-readable raw form within the benchmark list, the category-entry-combination is key-value-combined with a colon: \texttt{programming-model:cuda} (for \btProgMod{cuda}).

\begin{figure}
\begin{boxE}
\begin{description}
    \footnotesize
    \item[\btAppDomain{\!} \texttt{application-domain}] asc, astrophysics, automotive, bioinformatics, biology, cfd chemistry, climate, combustion, computer-vision, cosmology, cryptography, dft, engineering, finance, fusion, geoscience, hep, hydrodynamics, material-science, medical, molecular-dynamics, nuclear, physics, polymers, qcd, robotics, seismic, solarphysics, synthetic, thermodynamics
    \item[\btScale{\!} \texttt{benchmark-scale}] large-scale, multi-node, single-node, strong-scaling, sub-node, weak-scaling
    \item[\btComm{\!} \texttt{communication}] mpi, nccl, nvsmem, openshmem, rccl, shmem, upc, upc++
    \item[\btCommPerf{\!} \texttt{communication-performance-characteristics}] network-bandwidth-bound, network-bisection-bandwidth-bound, network-collectives, network-latency-bound, network-multi-threaded, network-nonblocking-collectives, network-onesided, network-point-to-point
    \item[\btCompPerf{\!} \texttt{compute-performance-characteristics}] atomics, high-branching, high-fp, i-o, mixed-precision, register-pressure, simd, vectorization
    \item[\btMemAcc{\!} \texttt{memory-access-characteristics}] high-memory-bandwidth, irregular-memory-access, large-memory-footprint, managed-memory, regular-memory-access
    \item[\btMesh{\!} \texttt{mesh-representation}] hamr, block-structured-grid, meshfree, multigrid, structured-grid, unstructured-grid
    \item[\btMethod{\!} \texttt{method-type}] ai, ale, compression, conjugate-gradient, dense-linear-algebra, deterministic, direct-solve, eulerian, explicit-differentiation, explicit-timestepping, fft, finite-difference, finite-element, finite-volume, full-assembly, graph, graph-traversal, high-order, hydrodynamics, implicit-differentiation, implicit-timestepping, lagrangian, lbm, low-order, math, montecarlo, nbody, no-method, ode, partial-assembly, particles, pde, rng, signal-processing, solver, sorting, sparse-linear-algebra, spatial-discretization, sph, task-parallelism, time-dependent, transport
    \item[\btProgLang{\!} \texttt{programming-language}] c, c++, fortran, java, julia, python, rust
    \item[\btProgMod{\!} \texttt{programming-model}] charm++, cuda, dpc++, futhark, hip, kokkos, oneapi, openacc, opencl, openmp, openmp-target, pstl, raja, rocm, sycl, tbb, thrust
\end{description}
\end{boxE}
\caption{Taxonomy overview with top-level \emph{categories} (printed in bold with added symbol) and each multiple \emph{entries}.}
\label{fig:tax}
\end{figure}

%% file: sections/5-benchmarks.tex
\section{Benchmark Survey}
\label{sec:benchmarks}

\newcommand{\inlineLicense}[1]{\raisebox{0.5ex}{\scalebox{0.8}{#1}}}

\supp{\autoref{table:benchmarks:short:c} presents a shortened overview of the surveyed benchmarks\footnote{\rev{Due to its size}, the table is inserted at the end of this article.}, augmenting the vast complete overview \autoref{table:benchmarks} available as supplemental material}. Over 180 individual benchmarks and 13 benchmark suites were collected. The \supp{main future-proof point of access is the online version of the} table at \href{https://fzj-jsc.github.io/benchmark-survey/}{\url{fzj-jsc.github.io/benchmark-survey/}} with features for dynamic filtering and sorting, as well as \supp{automatically} incorporating updates \supp{from the community}.

\supp{The shortened overview in \autoref{table:benchmarks:short:c} first lists suites of benchmarks with commentary about the suite and contained benchmarks, and then free-standing benchmarks with respective notes.}
\supp{The complete table \autoref{table:benchmarks} and the online version contain more details. Each} table entry consists of the most relevant information for a benchmark/suite:
\begin{description}
    \item[Name] The self-given name of each benchmark/suite is used for identification; when ambivalent, the more commonly used term was taken
    \item[Taxonomy Tags] Building up on the Benchmark Taxonomy of \autoref{sec:taxonomy}, each entry is characterized by a set of taxonomy tags. For suites, tags common to (nearly) all contained benchmarks are promoted to top-level tags of the suite itself and are not separately displayed for each individual benchmark.
    \item[License] As the scope of applicability of a benchmark is closely related to its license, an effort is made to collect the information here. For brevity, each license is shown in symbol form using the following keys:
        \inlineLicense{\licNone{}} No license; 
        \inlineLicense{\licFree{}} Free (no (clear) license specified, but freely available); 
        \inlineLicense{\licProp{}} Proprietary/Custom; 
        \href{https://spdx.org/licenses/MIT.html}{\inlineLicense{\licMit{}}} MIT; 
        \href{https://spdx.org/licenses/BSD-2-Clause.html}{\inlineLicense{\licBsdTwo{}}}/\href{https://spdx.org/licenses/BSD-3-Clause.html}{\inlineLicense{\licBsdThree{}}}/\href{https://spdx.org/licenses/BSD-4-Clause.html}{\inlineLicense{\licBsdFour{}}} BSD-2/3/4-Clause; 
        \href{https://spdx.org/licenses/GPL-2.0.html}{\inlineLicense{\licGplTwo{}}}/\href{https://spdx.org/licenses/GPL-3.0.html}{\inlineLicense{\licGplThree{}}} GPL-2.0/3.0; 
        \href{https://spdx.org/licenses/LGPL-2.1.html}{\inlineLicense{\licLgplTwoone{}}} LGPL-2.1 (logo abbreviates 2.1 to 2); 
        \href{https://spdx.org/licenses/Apache-2.0.html}{\inlineLicense{\licApacheTwo{}}} Apache-2.0; 
        \href{https://spdx.org/licenses/MPL-2.0.html}{\inlineLicense{\licMplTwo{}}} MPL-2.0; 
        \href{https://spdx.org/licenses/CC-BY-4.0.html}{\inlineLicense{\licCcByFour{}}} CC-BY-4.0.
    \item[Reference] For easy access, a URL for each benchmark is provided in \emph{\scalebox{0.8}{\faIcon{link}}link} form.
    \item[Notes] In the notes, comments and details about the benchmarks/suites are added, as well as names of benchmarks of suites-within-suites (HeCBench, SPEC ACCEL) and older suites (SPEC MPI, SPEC OMP).  If available, a scientific publication is added as reference at the end of the notes. But only a minority of benchmarks expose readily available publications.
\end{description}

The completeness of this metadata is highly dependent on the offered information at the source of the data; i.e. only if clear description is provided, it could be added as well-formed metadata. Especially, the taxonomy tags referring to workload profiles are highly reliant on the available data. In addition, the entries are augmented with the authors' knowledge of the workload, building up on their experiences in the field. \supp{The authors gladly welcome contributions by the community on GitHub to further extend this list.}

The collected data is available in machine-readable, raw YAML form on the accompanying \href{https://github.com/FZJ-JSC/benchmark-survey}{GitHub repository}~\citep{benchmarksyaml_link}. This allows for easy extension; and also for future developments building up on the data. The YAML schema uses the following keys for each entry: \texttt{name}, \texttt{tags}, \texttt{license}, \texttt{url}, \texttt{ref}, \texttt{notes}, \texttt{benchmarks}; \hyperref[lst:schema]{Listing~\ref{lst:schema}} gives an example. Except for \texttt{tags}, which expects a list of taxonomy tags (see \autoref{fig:tax}), every key is accompanied by a single value. The \texttt{benchmarks} key is only present for benchmark suites, and under it the individual benchmarks are listed with the same schema. Freestanding benchmarks outside of suites are summarized under a top-level key \texttt{benchmarks}; again, for each entry, the default schema follows.
\makeatletter
\let\@float@c@listing\@caption
\makeatother

\begin{listing}
    \begin{minted}[fontsize=\footnotesize,autogobble,breaklines]{yaml}
    benchmarks:
        hpl:
            name: HPL
            url: https://www.netlib.org/benchmark/hpl/
            license: BSD-4-Clause
            ref: 10.1145/141868.141871
            tags: [application-domain:synthetic, programming-language:c, math-libraries:blas, method-type:dense-linear-algebra]
            note: "High Performance Linpack"
    \end{minted}
    \caption{Example benchmark definition in YAML schema.}
    \label{lst:schema}
\end{listing}

%% file: sections/6-evaluation.tex
\section{Evaluation of Survey}
\label{sec:eval}

\subsection{Overview and Highlights}

In the \supp{complete} survey, 13 benchmark suites were recorded. They can be roughly sorted into three groups. References are omitted here, as they can be neatly found in \autoref{table:benchmarks} and \href{https://fzj-jsc.github.io/benchmark-survey/}{online}.

\begin{description}
    \item[System Procurements] These suites were created for large system procurements and used for evaluation of large installations. They are arguable the most thoroughly documented benchmarks surveyed, as they document the needs and requirements of whole user communities to system integrators. Nearly all benchmarks in this list are GPU-accelerated and combine application benchmarks (or mini-apps thereof) with synthetic benchmarks. 
    
    OLCF-6 is the suite for the next-generation supercomputer to be hosted at Oak Ridge National Lab, the RFP closed in autumn 2024. ATS-5 is the suite for the next supercomputer at Los Alamos National Lab, due to be installed in 2026/2027. NERSC-10 is the benchmark suite for the successor of the Perlmutter system at NERSC, to be deployed in 2026. The JUPITER Benchmark Suite~\cite{jupiter_ref} was used for the procurement of JUPITER, currently being built at Jülich Supercomputing Center. The CORAL-2 benchmark suite was used for the acquisitions of Frontier and El Capitan, hosted at Oak Ridge National Lab and Lawrence Livermore National Lab, respectively; the systems are already operational.
    \item[Research \& Community Collections] Grown out of endeavors of research institutes and partly supported by the community, a number of benchmark suites has emerged which focus on certain niches. For example RAJAPerf~\cite{10.1109/rajaperf2024}, which started as a suite to verify and showcase the RAJA programming model~\cite{RAJA-sc19}, but now incorporates other parallel programming models in comparative fashion; the focus is on typical parallel patterns and simple algorithms. RAJAPerf includes other  suites, for example LCALS and Polybench. HeCBench is similar and includes a vast amount of simple computational benchmarks in a variety of different GPU programming models. In the table, we only list the categories, as the suite contains more than 400 individual programs. Another well-known collection of benchmarks for heterogeneous machines is the Rodinia benchmark suite. It is frequently used in the community for many performance investigations. The future of Rodinia is unsure, as the benchmark currently appears unmaintained. The case is similar for UEABS (\emph{Unified European Applications Benchmark Suite}) by the European PRACE project. The suite captures many well-known applications with detailed execution instructions and input data, but appears not maintained anymore. The HPC Challenge benchmark suite combines many synthetic benchmarks, and also shares results on their website; but the project seems currently abandoned.
    \item[(Semi-)Commercial Offerings] The Standard Performance Evaluation Corporation, SPEC, provides different benchmark suites targeted for a variety of use-cases. Relevant for HPC are SPEChpc, offering a number of benchmarks in different workload sizes, SPEC ACCEL, last updated 2019, and focusing on different GPU benchmarks (with OpenCL, OpenMP, and OpenACC), and SPECaccel 2023, updating the previous suite and using OpenMP- and OpenACC-accelerated applications for benchmarking. Further benchmarks relevant to HPC exist. The benchmark setups are closed source and can be commercially acquired.
\end{description}

The choice for individual benchmarks is plentiful -- be it freestanding or integrated into a suite. Famous and well-used benchmarks are HPL, which is particular compute-intensive and used for the Top500 ranking, HPCG, a conjugate gradient benchmark with data access patterns seen in applications, STREAM, which measures memory bandwidth with simple data movement routines, BabelStream, a STREAM implementation for a variety of GPUs, the Ohio State University Micro-Benchmarks, which test communication libraries (for example MPI), MLPerf, one of the few AI/ML benchmark in the suite, or IOR, a tool to determine I/O performance also used for the IO500. 

Well-curated benchmarks are scarce, and the silent sunsetting of established suites is a loss for the field. It appears challenging to keep pace with the fast-moving update cycle in HPC\rev{, resulting in some benchmarks discontinued to various degrees}. The authors hope that with this work, not only an overview can be gotten, but also awareness raised for the trove of choice currently existing.

\subsection{Statistical Evaluation}

Albeit identification of characterizing aspects was at times challenging, many taxonomy tags could be attached to the benchmarks. The approach of using a YAML scheme for the taxonomy and the survey allows for some first, imperfect\footnote{Of course, this is not the core scope of the paper. We do not claim completeness in the added taxonomy tags, but merely survey. Also, we do not resolve into "suites-of-suites" (HeCBench).} evaluation of the state-of-the-practice. For the following numbers, taxonomy tags common to benchmarks in a suite have been, temporarily, added again to the benchmarks themselves. Categories can, of course, be present multiple times for individual benchmarks, for example if a benchmark is available in both OpenMP and CUDA.

Over 400 times a \scalebox{1.1}{\btProgMod{\!}}\,programming model was identified, with OpenMP being the most prominent parallelization choice (158 entries), followed by CUDA (94 times); OpenMP Target (48), OpenACC (43), and HIP (31) follow. Not claiming perfect representativeness, the results still appear to reflect the current trends in the field: CPU-focused benchmarks almost exclusively utilize OpenMP for CPU-parallelization and GPU benchmarks mostly use CUDA for GPU-parallelization. A \scalebox{1.1}{\btProgLang{\!}}\,programming languages could be identified 222 times, with C and C++ equally the most prominent ones (80 entries each). Fortran has 50 entries, Python only 10. Despite \rev{Fortran's importance} in HPC, the vast majority of benchmarks use C/C++. Surprising is the little count of Python benchmarks, being the driving force in many sub-domains of HPC. The most used \scalebox{1.1}{\btScale{\!}}\,benchmark scale is \emph{single-node} with 88 entries; \emph{multi-node} follows with 52 entries. A focus on intra-node tests can be seen, removing network effects and related implementation/evaluation complications. \scalebox{1.1}{\btAppDomain{\!}}\,Application information is available 122 times, with \emph{synthetic} being the front-runner (18), \emph{physics} (17), \emph{molecular dynamics} (MD, 17), and \emph{computational fluid} dynamics (CFD) and \emph{climate} following (both 12). A slight focus appears to be on synthetic benchmarks, determining intricate and specific hardware features inspired by actual application workloads. Physics, MD, CFD, climate research are traditional HPC use-cases, which are expectedly represented in the survey. 92 benchmarks were identified to use MPI for \scalebox{1.1}{\btComm{\!}}\,communication, with much distance to NCCL (5). MPI is the de-facto standard for communication in HPC, which can clearly be seen.

The taxonomy categories referring to more involved benchmark details are harder to identify and require thorough description or deeper knowledge of the benchmarks. Characteristics of \scalebox{1.1}{\btMemAcc{\!}}\,memory access could be identified 48 times, of \scalebox{1.1}{\btCommPerf{\!}}\,communication performance 40 times, and of \scalebox{1.1}{\btCompPerf{\!}}\,compute performance 28 times.

%% file: sections/7-conclusions.tex
\section{Conclusions}
\label{sec:conclusions}
Benchmarks are essential tools in the HPC field, enabling well-defined and comparable assessments of HPC systems. They characterize hardware features and software capabilities in an objective manner, support the advancement of the field, and are key for investment decisions.

In the presented work, we collect available HPC benchmarks and benchmark suites and survey them in a concise and comparable manner. \supp{An overview with strongly reduced level of detail is available in \autoref{table:benchmarks:short:c}. The much longer full survey, including characterization tags, is available as supplemental material} in \autoref{table:benchmarks} and \href{https://fzj-jsc.github.io/benchmark-survey/}{online \supp{(interactive)}}. 13 benchmark suites and over 180 benchmarks could be collected with detailed meta-data, like associated license, references, notes, and characterization. For the latter, we create a Benchmark Taxonomy (\autoref{fig:tax}) to describe different aspects of the individual benchmarks and suites and enable easy comparison. The raw data of the survey and the taxonomy are available for further extension and collaboration as open source software on GitHub, feeding directly into the \supp{interactive} website.

The benchmark suites are either created for large system procurement (like Frontier, El Capitan, or JUPITER), or come from parts of the community (like RAJAPerf). Without claiming representativeness, we attempt a first evaluation of the collected taxonomy tags and find, for example, OpenMP \rev{appears} to be by far the most-used programming model for parallelization on the CPU. On the GPU, CUDA is the main model.

In the future, we expect to further extend the Benchmark Taxonomy, adding, for example, further application domains, methods, or libraries\rev{; we also consider extending it by status-related information to cover functionality and topicality of benchmarks, since some defunct benchmarks were collected}. Although being thorough in our review of available benchmarks, we are certain that some benchmarks escaped our attention. We expect to extend the benchmark survey in the future to benchmarks not yet included in the current data; in hopes of help by the community online.

%% file: sections/8-ack.tex
\section{Acknowledgements}

The authors would like to thank Jens Domke for his thoughts relating to significant benchmarks for this work.

This work was performed under the auspices of the 
U.S.~Department of Energy by Lawrence Livermore National 
Laboratory under Contract DE-AC52-07NA27344
and was supported by the LLNL-LDRD Program under 
Project No. 24-SI-005
(LLNL-JRNL-2001672).

%% file: sections/9-bio.tex
\section{Author Biographies}

{\bf Dr. Andreas Herten} is a researcher at Jülich Supercomputing Centre (JSC) of Forschungszentrum Jülich (FZJ). He is a joint lead of the Novel System Architecture Design division, in which he heads the Accelerating Devices Lab. His research focuses on enabling applications for new hardware, especially GPUs and other accelerators, collaborating closely with vendors and users. He enjoys investigating programming models for performance and reproducible benchmarking of hardware-software ecosystems. Recently, he was responsible for the benchmarks used for assessment of JUPITER, the first European exascale supercomputer, where the thought for this survey was ignited.

Andreas is a physicist by training, receiving his doctorate from Ruhr-Uni Bochum/FZJ (institute IKP) for research on usage of highly-parallel algorithms on GPUs in hadron physics. He joined JSC in 2015 to continue enabling applications on GPUs. Since then, he is involved in a variety of third-party funded projects for HPC enablement and performance improvement, and participated in multiple HPC system procurements.

\vspace{1\baselineskip}

\noindent
{\bf Dr. Olga Pearce} is a computer scientist in the Center for Applied Scientific Computing at Lawrence Livermore National Laboratory. She created Benchpark, an open collaborative repository for reproducible specifications of HPC benchmarks and cross-site benchmarking environments, and Thicket, an open source toolkit for Exploratory Data Analysis (EDA) of parallel performance data. Olga leads benchmarking for Advanced Technology Systems, the Performance Analysis and Visualization for Exascale Project, and Performance Modeling in the Fractale SI. Her research interests include HPC architectures and simulations, parallel algorithms and programing models, system software, performance analysis and optimization.

Olga has been at LLNL since 2007. She received the NSF graduate fellowship in 2006, Lawrence Scholar Fellowship in 2009, and joined CASC as technical staff in 2014.  Olga received the LLNL Deputy Director’s Science \& Technology award (2015), and LLNL awards for developing the RAJA performance portability model (2018), porting and optimization of codes on LLNL’s first accelerated supercomputer (2019), developing GPU capabilities of the Next-Gen Multiphysics code (2021), response to the National Academies of Sciences RFI (2023), and acceptance of El Capitan (2022, 2024).  Olga helped create the SC Student Cluster Challenge in 2008, started a joint appointment at Texas A\&M University as the Associate Professor of Practice in the Computer Science and Engineering in 2021, and serves as a co-chair of the Salishan Conference on High-Speed Computing.  Olga received her Ph.D. in Computer Science from Texas A\&M University, and her B.S. in Computer Science and Mathematics (dual major) from Western Oregon University. 

\vspace{1\baselineskip}

\noindent
{\bf Dr. Filipe Guimar\~aes} is a Computational Physicist who earned his PhD in 2011. He joined FZJ in 2014 as a researcher in the PGI-1 institute, specializing in condensed matter physics.

In 2020, Filipe transitioned to JSC, where he became a member of the support group of the Application Optimization lab. His work focuses on providing advanced technical support and optimizing computational workflows for scientific applications. Filipe is also one of the developers of LLview, a powerful tool for monitoring and visualizing high-performance computing resources.

%% file: sections/benchmark_table-smallC.tex
\begin{longtblr}[
    caption={Overview Version of Benchmark Survey}, %
    entry = {Benchmark Overview Table},
    label = {table:benchmarks:short:c}
]{
        width   = \textwidth,
        colspec = {XX[6]},
        cells   = {font=\footnotesize},
        row{1}  = {font=\footnotesize\bfseries}, rowhead = 1,
        rowsep = 0.2pt
    }
    \emph{(Suite-)}Name & Description, Included Benchmarks (if Suite)  \\
\midrule
\textbf{ATS-5} & Suite for the procurement of the fifth version of the Advanced Technology System by LANL/NNSA with representative workloads of the centers.  
\\
& Contained benchmarks: Branson, AMG2023, Parthenon-VIBE, MLMD, UMT, MiniEM, SPARTA, LAMMPS
 \\
\textbf{benchpark} & The suite comes with re-implementations of well-known benchmarks/benchmark suites in a variety of programming models  
\\
& Contained benchmarks: AD, AMG2023, BabelStream, Branson, GENESIS, GPCNet, GROMACS, HPCG, HPL, IOR, Kripke, Laghos, LAMMPS, MDTest, MiniEM, OSU Micro-Benchmarks, Parthenon-VIBE, Phloem, Quicksilver, QWS, RAJAPerf, Remhos, SMB, STREAM
 \\
\textbf{CORAL-2} & The suite comes with re-implementations of well-known benchmarks/benchmark suites in a variety of programming models  
\\
& Contained benchmarks: HACC, NEKBONE, QMCPACK, LAMMPS, AMG, Kripke, Quicksilver, PENNANT, BDAS, DLS, CMB, STREAM, Stride, MLDL, IOR-MDTest-Simul-FTree, CLOMP, Pynamic, RAJAPerf, E3SM, VPIC, Laghos, ParallelIntegerSort, Havoq
 \\
\textbf{HeCBench} & A collection of simple heterogeneous computing benchmarks, aligned in categories  
\\
& Contained benchmarks: Automotive benchmarks, Bandwidth benchmarks, Bioinformatics benchmarks, Computer vision and image processing, Cryptography, Data compression and reduction, Data encoding, decoding, or verification, Finance, Geoscience, Graph and tree, Language and kernel features, Machine learning, Math, Random number generation, Search, Signal processing, Simulation, Sorting, Robotics
 \\
\textbf{HPC Challenge} & A benchmark suite of 7 benchmarks that measures a range of memory access patterns.  
\\
& Contained benchmarks: HPL, DGEMM, Stream, PTRANS, RandomAccess, FFT, b\_eff
 \\
\textbf{JUPITER Benchmark Suite} & The JUPITER Benchmark Suite is used for procurement of the JUPITER exascale system and consists of application and synthetic benchmarks. The vast majority of the benchmarks focus on GPU execution. The application benchmarks come in Base category (usually executing on 8 nodes, each 4 GPUs) and High-Scaling category (using between 500 and 650 nodes, each 4 GPUs).   
\\
& Contained benchmarks: Amber, Arbor, Chroma LQCD, GROMACS, ICON, JUQCS, nekRS, ParFlow, PIConGPU, Quantum ESPRESSO, SOMA, MMoCLIP, Megatron-LM, ResNet, DynQCD, NAStJA, Graph500, HPCG, HPL, IOR, LinkTest, OSU Micro-Benchmarks, STREAM, STREAM (GPU)
 \\
\textbf{NERSC-10} & This suite represents scientific workflows: simulation of complex scientific problems at high degrees of parallelism, large-scale analysis of experimental or observational data, machine learning, and the data-flow and control-flow needed to couple these activities in productive and efficient workflows.  
\\
& Contained benchmarks: Optical Properties of Materials workflow, Materials by Design Workflow, Metagenome Annotation Workflow, Lattice QCD Workflow, DeepCAM AI Workflow, TOAST3 Cosmic Microwave Background Workflow
 \\
\textbf{OLCF-6} & Suite for procurement of the next ORNL supercomputer (post-exascale), developed to capture the programming models, programming languages, numerical motifs, fields of science, and other modalities of investigation expected to make up the bulk of the usage upon deployment.  
\\
& Contained benchmarks: LAMMPS, M-PSDNS, MILC, QMCPACK, FORGE, Workflow
 \\
\textbf{RAJAPerf} & The RAJA performance suite is designed to explore performance of loop-based computational kernels of the sort found in HPC applications. In particular, it is used to compare runtime performance of kernels implemented using RAJA, and the same kernels implemented using standard or vendor-supported parallel programming models directly (such as CUDA and ROCm).  
\\
& Contained benchmarks: STREAM, PolyBench, LCALS, Halo Communication, Basic Patterns, Application Kernels, Algorithms
 \\
\textbf{Rodinia} & No updates since a long time; customized BSD-3-Clause license  
\\
& Contained benchmarks: Leukocyte, Heart Wall, MUMmerGPU, CFD Solver, LU Decomposition, HotSpot, Back Propagation, Needleman-Wunsch, Kmeans, Breadth-First Search, SRAD, Streamcluster, Particle Filter, PathFinder, Gaussian Elimination, k-Nearest Neighbors, LavaMD2, Myocyte, B+ Tree, GPUDWT, Hybrid Sort, Hotspot3D, Huffman
 \\
\textbf{SPEC ACCEL} & Commercial suite with distinct execution instructions, focusing on accelerators. Current version: v1.3. Individual benchmarks are combined here per category for brevity.  
\\
& Contained benchmarks: SPEC ACCEL\_OCL, SPEC ACCEL\_OACC, SPEC ACCEL\_OMP
 \\
\textbf{SPECaccel 2023} & Commercial suite with distinct execution instructions, focusing on accelerators. Update of the preceding SPEC ACCEL suite, extending selected benchmarks.  
\\
& Contained benchmarks: 403.stencil, 404.lbm, 450.md, 452.ep, 453.clvrleaf, 455.seismic, 456.spF, 457.spC, 459.miniGhost, 460.ilbdc, 463.swim, 470.bt
 \\
\textbf{SPEChpc} & SPEChpc collects its benchmarks into suites of different workload sizes: tiny, small, medium, large. Each size targets different number of tasks and higher memory usage.  
\\
& Contained benchmarks: LBM D2Q37, SOMA, Tealeaf, Cloverleaf, Minisweep, POT3D, SPH-EXA, HPGMG-FV, miniWeather
 \\
\textbf{UEABS} & The Unified European Application Benchmark Suite is a set of 13 application codes maintained by PRACE. The last release was in 2022, and it is probably not maintained anymore.  
\\
& Contained benchmarks: Alya, Code\_Saturne, CP2K, GADGET, GPAW, GROMACS, NAMD, NEMO, PFARM, QCD, Quantum ESPRESSO, SPECFEM3D, TensorFlow
 \\
HPL & High Performance Linpack  \\
HPCG & High Performance Conjugate Gradients is a complement to Linpack (HPL)  \\
PolyBench & A benchmark suite of 30 numerical computations with static control flow, extracted from operations in various application domains. Last commit in 2018.  \\
STREAM & Simple synthetic benchmark program that measures sustainable memory bandwidth (in MB/s) and the corresponding computation rate for simple vector kernels.  \\
BabelStream & STREAM in many models for many devices; also available: Julia, Rust, Scala, Java  \\
PTRANS & Matrix Transpose  \\
RandomAccess & GUPS (Giga Updates Per Second)  \\
FFT & 1d Discrete Fourier Transforms  \\
b\_eff & MPI benchmark for measuring effective accumulated bandwidth in a network; several message sizes, communication patterns and methods used.  \\
LCALS & Livermore Compiler Analysis Loop Suite, a collection of loop kernels based, in part, on historical Livermore Loops benchmarks. Original website currently not reachable.  \\
Graph500 & Linpack for graph problems.  Breadth First Search (BFS).  \\
Rodinia (Julia) & Julia-port of the Rodinia benchmarks, with single and multi-threaded implementations and Julia+CUDA; has the Rodinia license (customized BSD-3-Clause)  \\
Rodinia (DPC++) & DPC++-translation of Rodinia benchmarks (SYCL tag added for visibility)  \\
Rodinia (SYCL) & SYCL implementations of Rodinia benchmarks, currently deprecated (integrated into/maintained through HeCBench)  \\
OSU Micro-Benchmarks & A vast collection of network-related micro-benchmarks.  \\
SPEC MPI 2007 & MPI-targeted benchmark suite. Last update: 2009 (v2.0). It features the following benchmarks: 104.milc, 107.leslie3d, 113.GemsFDTD, 115.fds4, 121.pop2, 122.tachyon, 125.RAxML, 126.lammps, 127.wrf2, 128.GAPgeofem, 129.tera\_tf, 130.socorro, 132.zeusmp2, 137.lu, 142.dmilc, 143.dleslie, 145.lGemsFDTD, 147.l2wrf2  \\
GPCNet & Global Performance and Congestion Network Tests.  \\
SPEC OMP 2012 & OpenMP-focused benchmark, successor of SPEC OMP 2001 benchmark. Includes the following benchmarks: 350.md, 351.bwaves, 352.nab, 357.bt331, 358.botsalgn, 359.botsspar, 360.ilbdc, 362.fma3d, 363.swim, 367.imagick, 370.mgrid331, 371.applu331, 372.smithwa, 376.kdtree  \\
benchFFT & FFTW's FFT benchmark  \\
MLPerf HPC & MLPerf Training: HPC collects 4 HPC-related benchmarks: climate segmentation (DeepCAM), cosmology parameter prediction (CosmoFlow), quantum molecular modeling (DimeNet++), protein folding (AlphaFold2).  \\
IO500 & The IO500 benchmark captures user-experienced I/O performance with a variety of workloads.  \\
Fiber Miniapp & Suite of miniapps with: CCS QCD, FFVC, NICAM-DC, mVMC, NGS Analyzer, MODYLAS, NTChem, FFB. Different licenses. Last update: 2015.  \\
LULESH & Suite of proxy apps for 3D Lagrangian hydrodynamics on unstructured mesh. Current version is 2, but no update has happened in quite some time.  \\
\end{longtblr}

%% file: sections/benchmark_table-full.tex
\begin{longtblr}[
    caption={Full Benchmark Survey. Benchmark suites are listed first and indicated by leftmost entries with individual subordinated benchmark entries, connected by a dotted line. Per benchmark, tags, license (\emph{Lic.}), URL, reference (\emph{Ref.}), and notes are given; if each component is common to all benchmarks in a suite, the components are listed for the suite itself.}{Full Benchmark Survey}, %
    entry = {Benchmark Overview Table},
    label = {table:benchmarks}
]{
    width   = \textwidth,
    colspec = {X[0.01]|[dotted]X[0.8,l]X[2.2]ccX[3]},
    cells   = {font=\footnotesize},
    row{1}  = {font=\footnotesize\bfseries}, rowhead = 1,
    rowsep = 0.2pt
}
\SetCell[c=2]{l} \emph{(Suite-)}Name & & Tags & Lic. & URL & Notes, Ref. \\
\midrule
\SetCell[c=2]{l} ATS-5 & & \btComm{mpi} \btProgLang{c++} \btProgLang{fortran} \btProgMod{openmp} &
\licNone &
\href{https://lanl.github.io/benchmarks/}{\scalebox{0.8}{\faIcon{link}}}&
Suite for the procurement of the fifth version of the Advanced Technology System by LANL/NNSA with representative workloads of the centers. \\
& Branson & \btAppDomain{physics} \btScale{multi-node} \btMethod{montecarlo} \btMethod{particles} \btMethod{transport} \btProgMod{cuda} &
&
&
\\
& AMG2023 & \btAppDomain{asc} \btScale{large-scale} \btCommPerf{network-collectives} \btCommPerf{network-latency-bound} \btComm{mpi} \btCompPerf{high-branching} \btCompPerf{mixed-precision} \btMemAcc{high-memory-bandwidth} \btMemAcc{high-memory-latency} \btMemAcc{irregular-memory-access} \btMemAcc{large-memory-footprint} \btMesh{block-structured-grid} \btMethod{solver} \btMethod{sparse-linear-algebra} \btProgLang{c} \btProgMod{cuda} \btProgMod{openmp} &
&
&
AMG is a parallel algebraic multigrid solver for linear systems arising from problems on unstructured grids. \\
& Parthenon-VIBE & \btScale{multi-node} \btMesh{amr} \btMesh{block-structured-grid} \btMethod{finite-volume} \btProgMod{kokkos} &
&
&
Solves the Vector Inviscid Burgers' Equation on a block-AMR mesh. \\
& MLMD & \btAppDomain{molecular-dynamics} \btMath{pytorch} \btProgMod{kokkos} &
&
&
Benchmark for HIPPYNN driven kokkos-Lammps Molecular Dynamics.  Includes model training and inference. \\
& UMT & \btAppDomain{physics} \btScale{multi-node} \btMemAcc{irregular-memory-access} \btMesh{unstructured-grid} \btMethod{deterministic} \btMethod{particles} \btMethod{transport} \btProgLang{fortran} \btProgMod{openmp-target} &
&
&
\\
& MiniEM & \btAppDomain{electromagnetics} \btMath{tpetra} \btMesh{structured-grid} &
&
&
Proxy for EMPIRE \\
& SPARTA & \btAppDomain{chemistry} \btAppDomain{molecular-dynamics} \btScale{multi-node} \btMethod{direct-solve} \btMethod{montecarlo} &
&
&
Stochastic PArallel Rarefied-gas Time-accurate Analyzer. \\
& LAMMPS & \btAppDomain{molecular-dynamics} \btScale{large-scale} \btCommPerf{network-bandwidth-bound} \btCommPerf{network-latency-bound} \btComm{mpi} \btCompPerf{high-fp} \btMemAcc{high-memory-bandwidth} \btProgLang{c++} \btProgMod{cuda} \btProgMod{openmp} &
&
&
Short and long range interactions. \\
\SetCell[c=2]{l} benchpark & & \btProgLang{c} \btProgLang{c++} \btProgLang{python} \btProgMod{cuda} \btProgMod{openmp} \btProgMod{raja} &
\licApacheTwo &
\href{https://github.com/LLNL/benchpark}{\scalebox{0.8}{\faIcon{link}}}&
The suite comes with re-implementations of well-known benchmarks/benchmark suites in a variety of programming models \\
& AD & \btCompPerf{high-fp} \btMemAcc{high-memory-bandwidth} \btMethod{automatic-differentiation} \btMethod{compiler-transformation} \btProgLang{c} \btProgLang{c++} &
&
&
AD is an automatic differentiation benchmark using Enzyme, a tool that takes arbitrary existing code as LLVM IR and computes the derivative (and gradient) of that function. \\
& AMG2023 & \btAppDomain{asc} \btScale{large-scale} \btCommPerf{network-collectives} \btCommPerf{network-latency-bound} \btComm{mpi} \btCompPerf{high-branching} \btCompPerf{mixed-precision} \btMemAcc{high-memory-bandwidth} \btMemAcc{high-memory-latency} \btMemAcc{irregular-memory-access} \btMemAcc{large-memory-footprint} \btMesh{block-structured-grid} \btMethod{solver} \btMethod{sparse-linear-algebra} \btProgLang{c} \btProgMod{cuda} \btProgMod{openmp} &
&
&
AMG is a parallel algebraic multigrid solver for linear systems arising from problems on unstructured grids. \\
& BabelStream & \btAppDomain{synthetic} \btScale{single-node} \btMemAcc{high-memory-bandwidth} \btMemAcc{regular-memory-access} \btProgLang{c} \btProgMod{openmp} &
&
&
See dedicated entry. \\
& Branson & \btAppDomain{physics} \btScale{multi-node} \btMethod{montecarlo} \btMethod{particles} \btMethod{transport} \btProgMod{cuda} &
&
&
\\
& GENESIS & \btAppDomain{molecular-dynamics} \btScale{multi-node} \btComm{mpi} \btMethod{particles} \btProgMod{cuda} \btProgMod{openmp} &
&
&
GENESIS package contains two MD programs (atdyn and spdyn), trajectory analysis programs, and other useful tools. GENESIS (GENeralized-Ensemble SImulation System) has been developed mainly by Sugita group in RIKEN-CCS \\
& GPCNet & \btScale{large-scale} \btCommPerf{network-bandwidth-bound} \btCommPerf{network-bisection-bandwidth-bound} \btCommPerf{network-collectives} \btCommPerf{network-latency-bound} \btComm{mpi} \btProgLang{c} &
\licApacheTwo &
\href{https://github.com/netbench/GPCNET}{\scalebox{0.8}{\faIcon{link}}}&
See dedicated entry. \\
& GROMACS & \btAppDomain{molecular-dynamics} \btScale{large-scale} \btScale{multi-node} \btComm{mpi} \btProgLang{c++} \btProgMod{cuda} \btProgMod{sycl} &
\licLgplTwoone &
\href{https://www.gromacs.org}{\scalebox{0.8}{\faIcon{link}}}&
\\
& HPCG & \btScale{large-scale} \btComm{mpi} \btMethod{conjugate-gradient} \btMethod{solver} \btMethod{sparse-linear-algebra} \btProgLang{c++} \btProgMod{openmp} &
&
\href{https://www.hpcg-benchmark.org}{\scalebox{0.8}{\faIcon{link}}}&
See dedicated entry.  High Performance Conjugate Gradients is a complement to Linpack (HPL) \\
& HPL & &
&
&
See dedicated entry \\
& IOR & \btAppDomain{synthetic} \btScale{large-scale} \btComm{mpi} \btCompPerf{i-o} \btProgLang{c} &
&
&
File system metadata benchmark. Interleaved or Random I/O benchmark. Used for testing the performance of parallel filesystems and burst buffers using various interfaces and access patterns. \\
& Kripke & \btAppDomain{asc} \btScale{large-scale} \btCommPerf{network-latency-bound} \btComm{mpi} \btCompPerf{atomics} \btCompPerf{high-fp} \btCompPerf{mixed-precision} \btCompPerf{register-pressure} \btCompPerf{simd} \btCompPerf{vectorization} \btMemAcc{high-memory-bandwidth} \btMemAcc{high-memory-latency} \btMemAcc{large-memory-footprint} \btMemAcc{regular-memory-access} \btMesh{structured-grid} \btMethod{deterministic} \btMethod{sparse-linear-algebra} \btMethod{transport} \btProgLang{c++} \btProgMod{cuda} \btProgMod{openmp} \btProgMod{raja} &
&
\href{https://github.com/LLNL/Kripke.git}{\scalebox{0.8}{\faIcon{link}}}&
Kripke is a structured deterministic (Sn) transport using RAJA. It contains wavefront algorithms. \\
& Laghos & \btAppDomain{asc} \btScale{large-scale} \btComm{mpi} \btMath{mfem} \btMethod{conjugate-gradient} \btMethod{dense-linear-algebra} \btMethod{eulerian} \btMethod{explicit-timestepping} \btMethod{finite-element} \btMethod{high-order} \btMethod{hydrodynamics} \btMethod{lagrangian} \btMethod{time-dependent} \btProgLang{c++} \btProgMod{cuda} &
&
\href{https://github.com/CEED/Laghos}{\scalebox{0.8}{\faIcon{link}}}&
Laghos (LAGrangian High-Order Solver) is a miniapp that solves the time-dependent Euler equations of compressible gas dynamics in a moving Lagrangian frame using unstructured high-order finite element spatial discretization and explicit high-order time-stepping.  Laghos supports full assembly and partial assembly modes. \\
& LAMMPS & \btAppDomain{molecular-dynamics} \btScale{large-scale} \btCommPerf{network-bandwidth-bound} \btCommPerf{network-latency-bound} \btComm{mpi} \btCompPerf{high-fp} \btMemAcc{high-memory-bandwidth} \btProgLang{c++} \btProgMod{cuda} \btProgMod{openmp} &
&
&
Short and long range interactions. \\
& MDTest & \btAppDomain{synthetic} \btScale{large-scale} \btComm{mpi} \btCompPerf{i-o} \btProgLang{c} &
&
&
File system metadata benchmark.  \\
& MiniEM & \btAppDomain{electromagnetics} \btMath{tpetra} \btMesh{structured-grid} &
&
&
Proxy for EMPIRE \\
& OSU Micro-Benchmarks & \btAppDomain{synthetic} \btScale{large-scale} \btScale{multi-node} \btScale{single-node} \btCommPerf{network-bandwidth-bound} \btCommPerf{network-bisection-bandwidth-bound} \btCommPerf{network-collectives} \btCommPerf{network-latency-bound} \btCommPerf{network-multi-threaded} \btCommPerf{network-nonblocking-collectives} \btCommPerf{network-onesided} \btCommPerf{network-point-to-point} \btComm{mpi} \btComm{nccl} \btComm{openshmem} \btComm{rccl} \btComm{upc} \btComm{upc++} \btCompPerf{atomics} \btMemAcc{managed-memory} \btMemAcc{regular-memory-access} \btProgLang{c} \btProgLang{java} \btProgLang{python} \btProgMod{cuda} \btProgMod{openacc} \btProgMod{rocm} &
&
\href{https://mvapich.cse.ohio-state.edu/benchmarks/}{\scalebox{0.8}{\faIcon{link}}}&
See dedicated entry.  A vast collection of network-related micro-benchmarks. \\
& Parthenon-VIBE & \btScale{multi-node} \btMesh{amr} \btMesh{block-structured-grid} \btMethod{finite-volume} \btProgMod{kokkos} &
&
&
Solves the Vector Inviscid Burgers' Equation on a block-AMR mesh. \\
& Phloem & \btAppDomain{synthetic} \btScale{large-scale} \btScale{multi-node} \btScale{single-node} \btCommPerf{network-bandwidth-bound} \btCommPerf{network-bisection-bandwidth-bound} \btCommPerf{network-collectives} \btCommPerf{network-latency-bound} \btCommPerf{network-multi-threaded} \btCommPerf{network-nonblocking-collectives} \btCommPerf{network-onesided} \btCommPerf{network-point-to-point} \btComm{mpi} \btComm{nccl} \btComm{openshmem} \btComm{rccl} \btComm{upc} \btComm{upc++} \btCompPerf{atomics} \btMemAcc{managed-memory} \btMemAcc{regular-memory-access} \btProgLang{c} \btProgLang{java} \btProgLang{python} \btProgMod{cuda} \btProgMod{openacc} \btProgMod{rocm} &
&
\href{https://github.com/LLNL/phloem}{\scalebox{0.8}{\faIcon{link}}}&
This is a package containing three MPI benchmarks intended for use during system procurement \\
& Quicksilver & \btAppDomain{asc} \btScale{large-scale} \btCommPerf{network-latency-bound} \btComm{mpi} \btCompPerf{high-branching} \btCompPerf{register-pressure} \btMemAcc{high-memory-latency} \btMemAcc{irregular-memory-access} \btMemAcc{large-memory-footprint} \btMethod{montecarlo} \btMethod{transport} \btProgLang{c++} \btProgMod{cuda} \btProgMod{openmp} &
&
&
Monte Carlo transport benchmark with multi-group cross section lookups. One large kernel that is 1000’s of lines big. \\
& QWS & \btAppDomain{qcd} \btScale{weak-scaling} \btComm{mpi} \btProgLang{c++} &
&
\href{https://www.riken.jp/en/research/labs/r-ccs/field_theor/index.html}{\scalebox{0.8}{\faIcon{link}}}&
QWS benchmark for Lattice quantum chromodynamics simulation library for Fugaku and computers with wide SIMD. \\
& RAJAPerf & \btAppDomain{asc} \btScale{single-node} \btProgLang{c++} \btProgMod{cuda} \btProgMod{openmp} \btProgMod{raja} &
&
&
See dedicated entry. \\
& Remhos & \btAppDomain{asc} \btScale{large-scale} \btComm{mpi} \btMath{mfem} \btMethod{conjugate-gradient} \btMethod{dense-linear-algebra} \btMethod{eulerian} \btMethod{explicit-timestepping} \btMethod{finite-element} \btMethod{high-order} \btMethod{hydrodynamics} \btMethod{lagrangian} \btMethod{time-dependent} \btProgLang{c++} \btProgMod{cuda} &
&
\href{https://github.com/CEED/Remhos}{\scalebox{0.8}{\faIcon{link}}}&
Remhos (REMap High-Order Solver) is a miniapp that solves the pure advection equations that are used to perform monotonic and conservative discontinuous field interpolation (remap) as part of the Eulerian phase in Arbitrary Lagrangian Eulerian (ALE) simulations. \\
& SMB & \btAppDomain{synthetic} \btScale{large-scale} \btScale{multi-node} \btScale{single-node} \btCommPerf{network-bandwidth-bound} \btCommPerf{network-bisection-bandwidth-bound} \btCommPerf{network-collectives} \btCommPerf{network-latency-bound} \btCommPerf{network-multi-threaded} \btCommPerf{network-nonblocking-collectives} \btCommPerf{network-onesided} \btCommPerf{network-point-to-point} \btComm{mpi} \btComm{nccl} \btComm{openshmem} \btComm{rccl} \btComm{upc} \btComm{upc++} \btCompPerf{atomics} \btMemAcc{managed-memory} \btMemAcc{regular-memory-access} \btProgLang{c} \btProgLang{java} \btProgLang{python} \btProgMod{cuda} \btProgMod{openacc} \btProgMod{rocm} &
&
\href{https://github.com/sandialabs/SMB}{\scalebox{0.8}{\faIcon{link}}}&
A vast collection of network-related micro-benchmarks. \\
& STREAM & \btAppDomain{synthetic} \btScale{single-node} \btMemAcc{high-memory-bandwidth} \btMemAcc{regular-memory-access} \btProgLang{c} \btProgMod{openmp} &
&
&
See dedicated entry. Memory subsystem functionality and performance tests. \\
\SetCell[c=2]{l} CORAL-2 & & \btProgLang{c} \btProgLang{c++} \btProgLang{fortran} \btProgLang{python} \btProgMod{cuda} \btProgMod{openmp} &
\licNone &
\href{https://asc.llnl.gov/coral-2-benchmarks}{\scalebox{0.8}{\faIcon{link}}}&
The suite comes with re-implementations of well-known benchmarks/benchmark suites in a variety of programming models \\
& HACC & \btAppDomain{astrophysics} \btScale{large-scale} \btComm{mpi} \btCompPerf{high-fp} \btMethod{fft} \btMethod{nbody} \btProgLang{c++} \btProgMod{cuda} \btProgMod{openmp} &
&
&
\\
& NEKBONE & \btAppDomain{cfd} \btScale{large-scale} \btCommPerf{network-collectives} \btComm{mpi} \btCompPerf{high-fp} \btMethod{conjugate-gradient} \btProgLang{c} \btProgLang{fortran} \btProgMod{cuda} &
&
&
Nekbone is a mini-app derived from the Nek5000 CFD code which is a high order, incompressible Navier-Stokes CFD solver based on the spectral element method. \\
& QMCPACK & \btAppDomain{molecular-dynamics} \btScale{large-scale} \btComm{mpi} \btMemAcc{high-memory-bandwidth} \btMethod{montecarlo} \btProgLang{c} \btProgLang{c++} \btProgMod{cuda} \btProgMod{openmp} &
&
&
QMCPACK is a many-body ab initio quantum Monte Carlo code for computing the electronic structure of atoms, molecules, and solids. \\
& LAMMPS & \btAppDomain{molecular-dynamics} \btScale{large-scale} \btCommPerf{network-bandwidth-bound} \btCommPerf{network-latency-bound} \btComm{mpi} \btCompPerf{high-fp} \btMemAcc{high-memory-bandwidth} \btProgLang{c++} \btProgMod{cuda} \btProgMod{openmp} &
&
&
LAMMPS is a classical molecular dynamics code. \\
& AMG & \btAppDomain{asc} \btScale{large-scale} \btCommPerf{network-collectives} \btCommPerf{network-latency-bound} \btComm{mpi} \btCompPerf{high-branching} \btCompPerf{mixed-precision} \btMemAcc{high-memory-bandwidth} \btMemAcc{high-memory-latency} \btMemAcc{irregular-memory-access} \btMemAcc{large-memory-footprint} \btMesh{block-structured-grid} \btMethod{solver} \btMethod{sparse-linear-algebra} \btProgLang{c} \btProgMod{cuda} \btProgMod{openmp} &
&
&
AMG is a parallel algebraic multigrid solver for linear systems arising from problems on unstructured grids. \\
& Kripke & \btAppDomain{asc} \btScale{large-scale} \btCommPerf{network-latency-bound} \btComm{mpi} \btCompPerf{atomics} \btCompPerf{high-fp} \btCompPerf{mixed-precision} \btCompPerf{register-pressure} \btCompPerf{simd} \btCompPerf{vectorization} \btMemAcc{high-memory-bandwidth} \btMemAcc{high-memory-latency} \btMemAcc{large-memory-footprint} \btMemAcc{regular-memory-access} \btMesh{structured-grid} \btMethod{deterministic} \btMethod{sparse-linear-algebra} \btMethod{transport} \btProgLang{c++} \btProgMod{cuda} \btProgMod{openmp} \btProgMod{raja} &
&
&
Kripke is a structured deterministic (Sn) transport using RAJA. It contains wavefront algorithms. \\
& Quicksilver & \btAppDomain{asc} \btScale{large-scale} \btCommPerf{network-latency-bound} \btComm{mpi} \btCompPerf{high-branching} \btCompPerf{register-pressure} \btMemAcc{high-memory-latency} \btMemAcc{irregular-memory-access} \btMemAcc{large-memory-footprint} \btMethod{montecarlo} \btMethod{transport} \btProgLang{c++} \btProgMod{cuda} \btProgMod{openmp} &
&
&
Monte Carlo transport benchmark with multi-group cross section lookups. One large kernel that is 1000’s of lines big. \\
& PENNANT & \btAppDomain{asc} \btScale{large-scale} \btCommPerf{network-latency-bound} \btComm{mpi} \btMemAcc{high-memory-bandwidth} \btMemAcc{irregular-memory-access} \btMesh{unstructured-grid} \btMethod{hydrodynamics} \btProgLang{c++} \btProgMod{cuda} \btProgMod{openmp} &
&
&
PENNANT is hydrodynamics mini-app for unstructured meshes in 2D (arbitrary polygons). It makes heavy use of indirect addressing and irregular memory access patterns. \\
& BDAS & \btMethod{ai} &
&
&
The big data analytic suite contains the K-Means observation label, PCA, and SVM benchmarks. \\
& DLS & \btMethod{ai} &
&
&
The deep learning suite contains: Convolutional Neural Networks (CNNs), LSTM recurrent neural network (RNN) architecture, and CANDLE deep learning architectures relevant to problems in cancer. \\
& CMB & \btAppDomain{synthetic} \btScale{large-scale} \btCommPerf{network-bandwidth-bound} \btCommPerf{network-bisection-bandwidth-bound} \btCommPerf{network-collectives} \btCommPerf{network-latency-bound} \btComm{mpi} \btProgLang{c} &
&
&
Subsystem functionality and performance tests. Collection of independent MPI benchmarks to measure various aspects of MPI performance including interconnect messaging rate, latency, aggregate bandwidth, and collective latencies. \\
& STREAM & \btAppDomain{synthetic} \btScale{single-node} \btMemAcc{high-memory-bandwidth} \btMemAcc{regular-memory-access} \btProgLang{c} \btProgMod{openmp} &
&
&
See dedicated entry. Memory subsystem functionality and performance tests. \\
& Stride & \btAppDomain{synthetic} \btScale{single-node} \btMemAcc{high-memory-bandwidth} \btMemAcc{regular-memory-access} \btProgLang{c} \btProgMod{openmp} &
&
&
Memory subsystem functionality and performance tests. \\
& MLDL & \btCompPerf{mixed-precision} \btMethod{ai} \btMethod{dense-linear-algebra} \btMethod{fft} \btMethod{sparse-linear-algebra} &
&
&
Sparse and dense convolutions, FFT, double, single and half precision GEMM and other machine/deep learning math algorithms not included in other CORAL benchmark suites. \\
& IOR-MDTest-Simul-FTree & \btAppDomain{synthetic} &
&
&
File system metadata benchmark. Interleaved or Random I/O benchmark. Used for testing the performance of parallel filesystems and burst buffers using various interfaces and access patterns. \\
& CLOMP & \btAppDomain{synthetic} \btScale{single-node} \btProgLang{c} \btProgMod{openmp} &
&
&
Threading benchmark focused on performance of thread overheads evaluation. \\
& Pynamic & \btAppDomain{asc} \btScale{large-scale} \btComm{mpi} \btProgLang{c} \btProgLang{python} \btProgMod{cuda} \btProgMod{openmp} \btProgMod{raja} &
&
&
Subsystem functionality and performance test. Dummy application that closely models the footprint of an important Python-based multi-physics ASC code. \\
& RAJAPerf & \btAppDomain{asc} \btScale{single-node} \btProgLang{c++} \btProgMod{cuda} \btProgMod{openmp} \btProgMod{raja} &
&
&
See dedicated entry. \\
& E3SM & \btAppDomain{climate} \btScale{large-scale} \btComm{mpi} \btMemAcc{high-memory-bandwidth} \btProgLang{fortran} \btProgMod{cuda} \btProgMod{openmp} &
&
&
E3SM is a high-resolution climate simulation code for the entire Earth system, containing five major components for the atmosphere, ocean, land surface, sea ice, and land ice along with a coupler. \\
& VPIC & \btComm{mpi} \btMethod{particles} \btProgLang{c++} \btProgMod{openmp} &
&
&
Vector Particle-In-Cell is a general purpose particle-in-cell simulation code for modeling kinetic plasmas. \\
& Laghos & \btAppDomain{asc} \btScale{large-scale} \btComm{mpi} \btMath{mfem} \btMethod{conjugate-gradient} \btMethod{dense-linear-algebra} \btMethod{eulerian} \btMethod{explicit-timestepping} \btMethod{finite-element} \btMethod{high-order} \btMethod{hydrodynamics} \btMethod{lagrangian} \btMethod{time-dependent} \btProgLang{c++} \btProgMod{cuda} &
&
&
Laghos supports full assembly and partial assembly modes. \\
& ParallelIntegerSort & \btComm{mpi} \btProgLang{c} \btProgLang{fortran} &
&
&
Sorts a large number of 64-bit integers (from 0 to T) in parallel. The emphasis here is IO, all-to-all communication, and integer operations. \\
& Havoq & \btMemAcc{irregular-memory-access} \btMemAcc{large-memory-footprint} &
&
&
Massively parallel graph analysis algorithms for computing triangles, edges, vertices. Emphasizes load imbalance and irregular random memory accesses. \\
\SetCell[c=2]{l} HeCBench & & \btScale{single-node} \btProgLang{c++} \btProgMod{cuda} \btProgMod{hip} \btProgMod{openmp-target} \btProgMod{sycl} &
\licBsdThree &
\href{https://github.com/zjin-lcf/HeCBench}{\scalebox{0.8}{\faIcon{link}}}&
A collection of simple heterogeneous computing benchmarks, aligned in categories Ref.: \cite{hecbench_ref} \\
& Automotive benchmarks & \btAppDomain{automotive} &
&
&
Benchmark: daphne \\
& Bandwidth benchmarks & \btMemAcc{high-memory-bandwidth} \btMemAcc{regular-memory-access} &
&
&
Benchmarks: cmembench, babelstream, memcpy, memtest, randomAccess, shmembench, triad \\
& Bioinformatics benchmarks & \btAppDomain{biology} &
&
&
Benchmarks: all-pairs-distance, bsw, ccs, cm, deredundancy, diamond, epistasis, extend2, frna, fsm, ga, logan, minibude, minimap2, nbnxm, nw, pcc, prna, sa, snake \\
& Computer vision and image processing & \btAppDomain{computer-vision} &
&
&
A set of 49 benchmarks \\
& Cryptography & \btAppDomain{cryptography} &
&
&
Benchmarks: aes, bitcracker, chacha20, columnarSolver, ecdh, keccaktreehash, merkle, present \\
& Data compression and reduction & \btMethod{compression} &
&
&
A set of 25 benchmarks \\
& Data encoding, decoding, or verification & &
&
&
Benchmarks: ans, crc64, crs, entropy, jenkins-hash, ldpc, md5hash, murmurhash3 \\
& Finance & \btAppDomain{finance} &
&
&
Benchmarks: aop, black-scholes, binomial, bonds, libor \\
& Geoscience & \btAppDomain{geoscience} &
&
&
Benchmarks: aidw, coordinates, geodesic, hausdorff, haversine, stsg \\
& Graph and tree & \btMethod{graph} &
&
&
Benchmarks: cc, floydwarshall, floydwarshall2, gc, hbc, hungarian, mis, sssp, rsmt \\
& Language and kernel features & \btProgLang{c++} &
&
&
A set of 40 benchmarks \\
& Machine learning & \btMethod{ai} &
&
&
A set of 69 benchmarks \\
& Math & \btMethod{math} &
&
&
A set of 64 benchmarks \\
& Random number generation & \btMethod{rng} &
&
&
Benchmarks: mt, permutate, qrg, rng-wallace, sobol, urng \\
& Search & &
&
&
Benchmarks: bfs, bsearch, b+tree, grep, keogh, s8n, ss, sss, tsp \\
& Signal processing & \btMethod{fft} \btMethod{signal-processing} &
&
&
Benchmarks: extrema, fft, lombscargle, sosfil, zmddft \\
& Simulation & &
&
&
A set of 97 benchmarks \\
& Sorting & \btMethod{sorting} &
&
&
Benchmarks: bitonic-sort, hybridsort, is, merge, quicksort, radixsort, segsort, sort, sortKV, split, warpsort \\
& Robotics & \btAppDomain{robotics} &
&
&
Benchmarks: inversek2j, rodrigues \\
\SetCell[c=2]{l} HPC Challenge & & \btAppDomain{synthetic} \btMemAcc{high-memory-bandwidth} \btMemAcc{irregular-memory-access} \btMemAcc{large-memory-footprint} \btMemAcc{regular-memory-access} &
&
\href{https://hpcchallenge.org/hpcc}{\scalebox{0.8}{\faIcon{link}}}&
A benchmark suite of 7 benchmarks that measures a range of memory access patterns. Ref.: \cite{hpcc_ref} \\
& HPL & &
&
&
See dedicated entry \\
& DGEMM & &
&
&
See dedicated entry \\
& Stream & &
&
&
See dedicated entry \\
& PTRANS & &
&
&
See dedicated entry \\
& RandomAccess & &
&
&
See dedicated entry \\
& FFT & &
&
&
See dedicated entry \\
& b\_eff & &
&
&
See dedicated entry \\
\SetCell[c=2]{l} JUPITER Benchmark Suite & & &
\licMit &
\href{https://github.com/FZJ-JSC/jubench}{\scalebox{0.8}{\faIcon{link}}}&
The JUPITER Benchmark Suite is used for procurement of the JUPITER exascale system and consists of application and synthetic benchmarks. The vast majority of the benchmarks focus on GPU execution. The application benchmarks come in Base category (usually executing on 8 nodes, each 4 GPUs) and High-Scaling category (using between 500 and 650 nodes, each 4 GPUs).  Ref.: \cite{jupiter_ref} \\
& Amber & \btAppDomain{molecular-dynamics} \btScale{single-node} \btComm{mpi} \btComm{nccl} \btProgLang{fortran} \btProgMod{cuda} &
\licProp &
\href{https://github.com/FZJ-JSC/jubench-amber/}{\scalebox{0.8}{\faIcon{link}}}&
\\
& Arbor & \btAppDomain{neuroscience} \btScale{large-scale} \btScale{multi-node} \btScale{weak-scaling} \btComm{mpi} \btProgLang{c++} \btProgMod{cuda} \btProgMod{hip} &
\licBsdThree &
\href{https://github.com/FZJ-JSC/jubench-arbor}{\scalebox{0.8}{\faIcon{link}}}&
The High-Scaling variant of the benchmark targets 642 nodes and comes in four memory configurations \\
& Chroma LQCD & \btAppDomain{hep} \btAppDomain{physics} \btScale{large-scale} \btScale{multi-node} \btScale{weak-scaling} \btComm{mpi} \btMath{quda} \btProgLang{c++} \btProgMod{cuda} \btProgMod{hip} &
\licProp &
\href{https://github.com/FZJ-JSC/jubench-chroma-qcd}{\scalebox{0.8}{\faIcon{link}}}&
The High-Scaling variant of the benchmark targets 512 nodes and comes in three memory configurations; License: JLab License \\
& GROMACS & \btAppDomain{molecular-dynamics} \btScale{large-scale} \btScale{multi-node} \btComm{mpi} \btProgLang{c++} \btProgMod{cuda} \btProgMod{sycl} &
\licLgplTwoone &
\href{https://github.com/FZJ-JSC/jubench-gromacs}{\scalebox{0.8}{\faIcon{link}}}&
\\
& ICON & \btAppDomain{climate} \btScale{large-scale} \btScale{multi-node} \btComm{mpi} \btProgLang{c} \btProgLang{fortran} \btProgMod{cuda} \btProgMod{hip} \btProgMod{openacc} &
\licBsdThree &
\href{https://github.com/FZJ-JSC/jubench-icon}{\scalebox{0.8}{\faIcon{link}}}&
\\
& JUQCS & \btAppDomain{quantum-computing} \btScale{large-scale} \btScale{multi-node} \btScale{weak-scaling} \btComm{mpi} \btProgLang{fortran} \btProgMod{cuda} \btProgMod{openmp} &
\licNone &
\href{https://github.com/FZJ-JSC/jubench-juqcs}{\scalebox{0.8}{\faIcon{link}}}&
The High-Scaling variant of the benchmark targets 512 nodes and comes in two memory configurations. In addition, a version is provided with runs at the same time on a CPU partition and a GPU partition (Modular Supercomputing Architecture). \\
& nekRS & \btAppDomain{cfd} \btScale{large-scale} \btScale{multi-node} \btScale{weak-scaling} \btComm{mpi} \btProgLang{c} \btProgLang{c++} \btProgMod{cuda} \btProgMod{hip} \btProgMod{occa} \btProgMod{sycl} &
\licBsdThree &
\href{https://github.com/FZJ-JSC/jubench-nekrs}{\scalebox{0.8}{\faIcon{link}}}&
The High-Scaling variant of the benchmark targets 642 nodes and comes in three memory configurations \\
& ParFlow & \btAppDomain{climate} \btAppDomain{hydrodynamics} \btScale{multi-node} \btComm{mpi} \btMath{hypre} \btProgLang{c} \btProgMod{cuda} \btProgMod{hip} &
\licLgplTwoone &
\href{https://github.com/FZJ-JSC/jubench-parflow}{\scalebox{0.8}{\faIcon{link}}}&
\\
& PIConGPU & \btAppDomain{physics} \btAppDomain{plasma-physics} \btScale{large-scale} \btScale{multi-node} \btScale{weak-scaling} \btComm{mpi} \btProgLang{c++} \btProgMod{alpaka} \btProgMod{cuda} \btProgMod{hip} &
\licGplThree &
\href{https://github.com/FZJ-JSC/jubench-picongpu}{\scalebox{0.8}{\faIcon{link}}}&
The High-Scaling variant of the benchmark targets 640 nodes and comes in three memory configurations \\
& Quantum ESPRESSO & \btAppDomain{dft} \btAppDomain{material-science} \btAppDomain{physics} \btScale{multi-node} \btComm{mpi} \btMath{elpa} \btProgLang{fortran} \btProgMod{cuf} \btProgMod{openacc} &
\licGplThree &
\href{https://github.com/FZJ-JSC/jubench-qe}{\scalebox{0.8}{\faIcon{link}}}&
\\
& SOMA & \btAppDomain{material-science} \btAppDomain{polymers} \btScale{multi-node} \btComm{mpi} \btComm{nccl} \btProgLang{c} \btProgMod{openacc} &
\licGplThree &
\href{https://github.com/FZJ-JSC/jubench-soma}{\scalebox{0.8}{\faIcon{link}}}&
\\
& MMoCLIP & \btAppDomain{computer-vision} \btAppDomain{llm} \btScale{multi-node} \btComm{mpi} \btComm{nccl} \btMath{pytorch} \btMethod{ai} \btProgLang{python} \btProgMod{cuda} \btProgMod{rocm} &
\licMit &
\href{https://github.com/FZJ-JSC/jubench-mmoclip}{\scalebox{0.8}{\faIcon{link}}}&
\\
& Megatron-LM & \btAppDomain{llm} \btScale{large-scale} \btScale{multi-node} \btComm{mpi} \btComm{nccl} \btMath{pytorch} \btMethod{ai} \btMethod{transformer} \btProgLang{python} \btProgMod{cuda} \btProgMod{rocm} &
\licBsdThree &
\href{https://github.com/FZJ-JSC/jubench-megatron-lm}{\scalebox{0.8}{\faIcon{link}}}&
\\
& ResNet & \btAppDomain{computer-vision} \btScale{multi-node} \btComm{mpi} \btMath{tensorflow} \btMethod{ai} \btProgLang{python} \btProgMod{cuda} \btProgMod{rocm} &
\licApacheTwo &
\href{https://github.com/FZJ-JSC/jubench-resnet}{\scalebox{0.8}{\faIcon{link}}}&
\\
& DynQCD & \btAppDomain{hep} \btAppDomain{physics} \btScale{multi-node} \btComm{mpi} \btProgLang{c} \btProgMod{openmp} &
\licNone &
\href{https://github.com/FZJ-JSC/jubench-dynqcd}{\scalebox{0.8}{\faIcon{link}}}&
\\
& NAStJA & \btAppDomain{biology} \btScale{multi-node} \btComm{mpi} \btProgLang{c++} &
\licMplTwo &
\href{https://github.com/FZJ-JSC/jubench-nastja}{\scalebox{0.8}{\faIcon{link}}}&
\\
& Graph500 & \btAppDomain{synthetic} \btScale{large-scale} \btScale{multi-node} \btComm{mpi} \btMethod{graph-traversal} \btProgLang{c} &
\licMit &
\href{https://github.com/FZJ-JSC/jubench-graph500}{\scalebox{0.8}{\faIcon{link}}}&
\\
& HPCG & \btAppDomain{synthetic} \btScale{large-scale} \btScale{multi-node} \btScale{single-node} \btComm{mpi} \btMethod{conjugate-gradient} \btProgLang{c++} \btProgMod{cuda} \btProgMod{hip} \btProgMod{openmp} &
\licBsdThree &
\href{https://github.com/FZJ-JSC/jubench-hpcg}{\scalebox{0.8}{\faIcon{link}}}&
\\
& HPL & \btAppDomain{synthetic} \btScale{large-scale} \btScale{multi-node} \btScale{single-node} \btComm{mpi} \btProgLang{c} \btProgMod{cuda} \btProgMod{hip} \btProgMod{openmp} &
\licBsdFour &
\href{https://github.com/FZJ-JSC/jubench-hpl}{\scalebox{0.8}{\faIcon{link}}}&
\\
& IOR & \btAppDomain{synthetic} \btScale{large-scale} \btScale{multi-node} \btScale{single-node} \btComm{mpi} \btCompPerf{i-o} \btProgLang{c} &
\licGplTwo &
\href{https://github.com/FZJ-JSC/jubench-ior}{\scalebox{0.8}{\faIcon{link}}}&
\\
& LinkTest & \btAppDomain{synthetic} \btScale{large-scale} \btScale{multi-node} \btCommPerf{network-point-to-point} \btComm{mpi} \btProgLang{c} &
\licBsdFour &
\href{https://github.com/FZJ-JSC/jubench-linktest}{\scalebox{0.8}{\faIcon{link}}}&
\\
& OSU Micro-Benchmarks & \btAppDomain{synthetic} \btScale{multi-node} \btScale{single-node} \btComm{mpi} \btProgLang{c} \btProgMod{cuda} \btProgMod{openacc} \btProgMod{rocm} &
\licBsdThree &
\href{https://github.com/FZJ-JSC/jubench-osu}{\scalebox{0.8}{\faIcon{link}}}&
Also see the dedicated entry. \\
& STREAM & \btAppDomain{synthetic} \btScale{single-node} \btMemAcc{regular-memory-access} \btProgLang{c} &
\licProp &
\href{https://github.com/FZJ-JSC/jubench-stream}{\scalebox{0.8}{\faIcon{link}}}&
Also see the dedicated entry. \\
& STREAM (GPU) & \btAppDomain{synthetic} \btScale{single-node} \btMemAcc{regular-memory-access} \btProgLang{c++} \btProgMod{cuda} &
\licFree &
\href{https://github.com/FZJ-JSC/jubench-stream-gpu}{\scalebox{0.8}{\faIcon{link}}}&
\\
\SetCell[c=2]{l} NERSC-10 & & \btComm{mpi} \btProgLang{c++} \btProgLang{fortran} \btProgMod{openmp} &
\licNone &
\href{https://www.nersc.gov/systems/nersc-10/benchmarks/}{\scalebox{0.8}{\faIcon{link}}}&
This suite represents scientific workflows: simulation of complex scientific problems at high degrees of parallelism, large-scale analysis of experimental or observational data, machine learning, and the data-flow and control-flow needed to couple these activities in productive and efficient workflows. \\
& Optical Properties of Materials workflow & \btAppDomain{dft} \btProgLang{c} \btProgLang{c++} \btProgLang{fortran} \btProgMod{openacc} \btProgMod{openmp} \btProgMod{openmp-target} &
&
\href{https://berkeleygw.org}{\scalebox{0.8}{\faIcon{link}}}&
BerkeleyGW's Epsilon module computes the material's dielectric function; the Sigma module computes the electronic self energy. \\
& Materials by Design Workflow & \btAppDomain{molecular-dynamics} &
&
&
EXAALT (ECP project), LAMMPS MD engine \\
& Metagenome Annotation Workflow & \btAppDomain{biology} &
&
&
Based on the computation workflow of the Joint Genome Institute's Integrated Microbial Genomes project. \\
& Lattice QCD Workflow & \btAppDomain{qcd} &
&
&
MILC Collaboration \\
& DeepCAM AI Workflow & \btAppDomain{climate} &
&
&
Trains a deep learning model to identify extreme weather phenomena (tropical cyclones, atmospheric rivers) in CAM5 climate simulation data. \\
& TOAST3 Cosmic Microwave Background Workflow & \btAppDomain{cosmology} &
&
&
Time Ordered Astrophysics Scalable Tools \\
\SetCell[c=2]{l} OLCF-6 & & \btComm{mpi} \btProgLang{c++} \btProgLang{fortran} \btProgMod{openmp} &
\licNone &
\href{https://www.olcf.ornl.gov/draft-olcf-6-technical-requirements/benchmarks}{\scalebox{0.8}{\faIcon{link}}}&
Suite for procurement of the next ORNL supercomputer (post-exascale), developed to capture the programming models, programming languages, numerical motifs, fields of science, and other modalities of investigation expected to make up the bulk of the usage upon deployment. \\
& LAMMPS & \btAppDomain{molecular-dynamics} \btScale{large-scale} \btCommPerf{network-bandwidth-bound} \btCommPerf{network-latency-bound} \btComm{mpi} \btCompPerf{high-fp} \btMemAcc{high-memory-bandwidth} \btProgLang{c++} \btProgMod{cuda} \btProgMod{openmp} &
&
&
Short and long range interactions. \\
& M-PSDNS & \btAppDomain{cfd} \btScale{large-scale} \btCommPerf{network-bandwidth-bound} \btCommPerf{network-bisection-bandwidth-bound} \btCommPerf{network-latency-bound} \btCommPerf{network-point-to-point} \btMethod{direct-solve} \btMethod{fft} \btProgMod{hip} &
&
&
Minimalist version of PSDNS. \\
& MILC & \btAppDomain{qcd} \btScale{large-scale} \btScale{strong-scaling} \btCommPerf{network-bandwidth-bound} \btCommPerf{network-latency-bound} \btMath{quda} &
&
&
Generation on su3-rhmd-hisq problem from the MIMD Lattice Collaboration (MILC) code. \\
& QMCPACK & \btAppDomain{molecular-dynamics} \btScale{large-scale} \btComm{mpi} \btMemAcc{high-memory-bandwidth} \btMethod{montecarlo} \btProgLang{c} \btProgLang{c++} \btProgMod{cuda} \btProgMod{openmp} &
&
\href{https://www.qmcpack.org}{\scalebox{0.8}{\faIcon{link}}}&
QMCPACK is a many-body ab initio quantum Monte Carlo code for computing the electronic structure of atoms, molecules, and solids. \\
& FORGE & \btCompPerf{low-precision} \btCompPerf{pipeline-parallelism} \btCompPerf{tensor} \btMethod{ai} \btMethod{ai-training} \btMethod{gpt} \btMethod{llm} \btMethod{transformer} &
&
&
For scientific downstream tasks, such as domain-subject classification. \\
& Workflow & \btMethod{ai} \btMethod{ai-training} \btMethod{particles} \btMethod{tpt} \btMethod{transformer} &
&
&
This benchmark aims to evaluate the efficiency of a HPC system in handling dynamic workloads from data streams. \\
\SetCell[c=2]{l} RAJAPerf & & \btScale{single-node} \btProgLang{c++} \btProgMod{cuda} \btProgMod{hip} \btProgMod{openmp} \btProgMod{openmp-target} \btProgMod{raja} \btProgMod{sycl} &
\licBsdThree &
\href{https://github.com/LLNL/RAJAPerf/}{\scalebox{0.8}{\faIcon{link}}}&
The RAJA performance suite is designed to explore performance of loop-based computational kernels of the sort found in HPC applications. In particular, it is used to compare runtime performance of kernels implemented using RAJA, and the same kernels implemented using standard or vendor-supported parallel programming models directly (such as CUDA and ROCm). Ref.: \cite{rajaperf_ref} \\
& STREAM & \btMemAcc{regular-memory-access} \btProgMod{kokkos} &
&
&
See dedicated entry \\
& PolyBench & &
&
&
See dedicated entry \\
& LCALS & \btMemAcc{regular-memory-access} \btProgMod{kokkos} &
&
&
See dedicated entry \\
& Halo Communication & &
&
&
\\
& Basic Patterns & \btProgMod{kokkos} &
&
&
\\
& Application Kernels & &
&
&
Variety of kernels extracted from real-world applications \\
& Algorithms & &
&
&
Algorithm primites (Atomic, Memcopy, Memset, Reduce, Scan, Sort, Sort-Pairs) \\
\SetCell[c=2]{l} Rodinia & & \btScale{single-node} \btProgLang{c} \btProgMod{cuda} \btProgMod{opencl} \btProgMod{openmp} &
\licFree &
\href{https://www.cs.virginia.edu/rodinia/doku.php?id=start}{\scalebox{0.8}{\faIcon{link}}}&
No updates since a long time; customized BSD-3-Clause license Ref.: \cite{rodinia_ref} \\
& Leukocyte & \btAppDomain{medical} \btMesh{structured-grid} &
&
\href{https://www.cs.virginia.edu/rodinia/doku.php?id=leukocyte}{\scalebox{0.8}{\faIcon{link}}}&
\\
& Heart Wall & \btAppDomain{medical} \btMesh{structured-grid} &
&
\href{https://www.cs.virginia.edu/rodinia/doku.php?id=heart_wall}{\scalebox{0.8}{\faIcon{link}}}&
\\
& MUMmerGPU & \btAppDomain{bioinformatics} \btMethod{graph-traversal} &
&
\href{https://www.cs.virginia.edu/rodinia/doku.php?id=mummergpu}{\scalebox{0.8}{\faIcon{link}}}&
\\
& CFD Solver & \btAppDomain{cfd} \btMesh{unstructured-grid} \btMethod{finite-volume} &
&
\href{https://www.cs.virginia.edu/rodinia/doku.php?id=cfd_solver}{\scalebox{0.8}{\faIcon{link}}}&
\\
& LU Decomposition & \btMethod{dense-linear-algebra} &
&
\href{https://www.cs.virginia.edu/rodinia/doku.php?id=lu_decomposition}{\scalebox{0.8}{\faIcon{link}}}&
\\
& HotSpot & \btAppDomain{physics} \btMesh{structured-grid} \btMethod{ode} &
&
\href{https://www.cs.virginia.edu/rodinia/doku.php?id=hotspot}{\scalebox{0.8}{\faIcon{link}}}&
\\
& Back Propagation & \btMesh{unstructured-grid} \btMethod{ai} &
&
\href{https://www.cs.virginia.edu/rodinia/doku.php?id=back_propagation}{\scalebox{0.8}{\faIcon{link}}}&
\\
& Needleman-Wunsch & \btAppDomain{bioinformatics} &
&
\href{https://www.cs.virginia.edu/rodinia/doku.php?id=needleman-wunsch}{\scalebox{0.8}{\faIcon{link}}}&
\\
& Kmeans & \btAppDomain{data-mining} \btMethod{clustering} \btMethod{dense-linear-algebra} &
&
\href{https://www.cs.virginia.edu/rodinia/doku.php?id=kmeans}{\scalebox{0.8}{\faIcon{link}}}&
\\
& Breadth-First Search & \btMethod{graph-traversal} &
&
\href{https://www.cs.virginia.edu/rodinia/doku.php?id=graph_traversal}{\scalebox{0.8}{\faIcon{link}}}&
\\
& SRAD & \btAppDomain{computer-vision} \btMesh{structured-grid} \btMethod{pde} &
&
\href{https://www.cs.virginia.edu/rodinia/doku.php?id=srad}{\scalebox{0.8}{\faIcon{link}}}&
\\
& Streamcluster & \btAppDomain{data-mining} \btMethod{clustering} \btMethod{dense-linear-algebra} &
&
\href{https://www.cs.virginia.edu/rodinia/doku.php?id=streamcluster}{\scalebox{0.8}{\faIcon{link}}}&
\\
& Particle Filter & \btAppDomain{medical} \btMesh{structured-grid} &
&
\href{https://www.cs.virginia.edu/rodinia/doku.php?id=particle_filter}{\scalebox{0.8}{\faIcon{link}}}&
\\
& PathFinder & \btMesh{structured-grid} &
&
\href{https://www.cs.virginia.edu/rodinia/doku.php?id=pathfinder}{\scalebox{0.8}{\faIcon{link}}}&
\\
& Gaussian Elimination & \btMethod{dense-linear-algebra} &
&
\href{https://www.cs.virginia.edu/rodinia/doku.php?id=gaussian_elimination}{\scalebox{0.8}{\faIcon{link}}}&
\\
& k-Nearest Neighbors & \btMethod{dense-linear-algebra} &
&
\href{https://www.cs.virginia.edu/rodinia/doku.php?id=k-nearest_neighbors}{\scalebox{0.8}{\faIcon{link}}}&
\\
& LavaMD2 & \btAppDomain{molecular-dynamics} \btMethod{paricles} &
&
\href{https://www.cs.virginia.edu/rodinia/doku.php?id=lavamd2}{\scalebox{0.8}{\faIcon{link}}}&
\\
& Myocyte & \btAppDomain{biology} \btMesh{structured-grid} \btMethod{ode} \btMethod{solver} &
&
\href{https://www.cs.virginia.edu/rodinia/doku.php?id=myocyte}{\scalebox{0.8}{\faIcon{link}}}&
\\
& B+ Tree & \btMethod{graph-traversal} &
&
\href{https://www.cs.virginia.edu/rodinia/doku.php?id=b_tree}{\scalebox{0.8}{\faIcon{link}}}&
\\
& GPUDWT & \btMethod{spectral} &
&
\href{https://www.cs.virginia.edu/rodinia/doku.php?id=gpudwt}{\scalebox{0.8}{\faIcon{link}}}&
\\
& Hybrid Sort & \btCompPerf{atomics} \btMethod{sorting} &
&
\href{https://www.cs.virginia.edu/rodinia/doku.php?id=hybrid_sort}{\scalebox{0.8}{\faIcon{link}}}&
\\
& Hotspot3D & \btAppDomain{physics} \btMesh{structured-grid} &
&
\href{https://www.cs.virginia.edu/rodinia/doku.php?id=hotspot3d}{\scalebox{0.8}{\faIcon{link}}}&
\\
& Huffman & \btMethod{state-machine} &
&
\href{https://www.cs.virginia.edu/rodinia/doku.php?id=huffman}{\scalebox{0.8}{\faIcon{link}}}&
\\
\SetCell[c=2]{l} SPEC ACCEL & & \btScale{single-node} \btScale{sub-node} \btProgLang{c} \btProgLang{c++} \btProgLang{fortran} \btProgMod{openacc} \btProgMod{opencl} \btProgMod{openmp-target} &
\licProp &
\href{https://www.spec.org/accel/}{\scalebox{0.8}{\faIcon{link}}}&
Commercial suite with distinct execution instructions, focusing on accelerators. Current version: v1.3. Individual benchmarks are combined here per category for brevity. \\
& SPEC ACCEL\_OCL & \btProgLang{c} \btProgLang{c++} \btProgMod{opencl} &
&
&
19 OpenCL-enabled benchmarks: 101.tpacf, 103.stencil, 104.lbm, 110.fft, 112.spmv, 114.mriq, 116.histo, 117.bfs, 118.cutcp, 120.kmeans, 121.lavamd, 122.cfd, 123.nw, 124.hotspot, 125.lud, 126.ge, 127.srad, 128.heartwall, 140.bplustree \\
& SPEC ACCEL\_OACC & \btProgLang{c} \btProgLang{fortran} \btProgMod{openacc} &
&
&
15 OpenACC-enabled benchmarks: 303.ostencil, 304.olbm, 314.omriq, 350.md, 351.palm, 352.ep, 353.clvrleaf, 354.cg, 355.seismic, 356.sp, 357.csp, 359.miniGhost, 360.ilbdc, 363.swim, 370.bt \\
& SPEC ACCEL\_OMP & \btProgLang{c} \btProgLang{fortran} \btProgMod{openmp} \btProgMod{openmp-target} &
&
&
15 OpenMP-enabled benchmarks: 503.postencil, 504.polbm, 514.pomriq, 550.pmd, 551.ppalm, 552.pep, 553.pclvrleaf, 554.pcg, 555.pseismic, 556.psp, 557.pcsp, 559.pmniGhost, 560.pilbdc, 563.pswim, 570.pbt \\
\SetCell[c=2]{l} SPECaccel 2023 & & \btScale{single-node} \btScale{sub-node} \btProgMod{openacc} \btProgMod{openmp} \btProgMod{openmp-target} &
\licProp &
\href{https://www.spec.org/accel2023/}{\scalebox{0.8}{\faIcon{link}}}&
Commercial suite with distinct execution instructions, focusing on accelerators. Update of the preceding SPEC ACCEL suite, extending selected benchmarks. \\
& 403.stencil & \btAppDomain{thermodynamics} \btMesh{structured-grid} \btMethod{pde} \btMethod{solver} \btProgLang{c} &
&
\href{https://www.spec.org/accel2023/Docs/benchmarks/403.stencil.html}{\scalebox{0.8}{\faIcon{link}}}&
\\
& 404.lbm & \btAppDomain{cfd} \btAppDomain{material-science} \btMethod{lbm} \btProgLang{c} &
&
\href{https://www.spec.org/accel2023/Docs/benchmarks/404.lbm.html}{\scalebox{0.8}{\faIcon{link}}}&
\\
& 450.md & \btAppDomain{molecular-dynamics} \btProgLang{fortran} &
&
\href{https://www.spec.org/accel2023/Docs/benchmarks/450.md.html}{\scalebox{0.8}{\faIcon{link}}}&
\\
& 452.ep & \btProgLang{c} &
&
\href{https://www.spec.org/accel2023/Docs/benchmarks/452.ep.html}{\scalebox{0.8}{\faIcon{link}}}&
\\
& 453.clvrleaf & \btAppDomain{climate} \btAppDomain{hydrodynamics} \btMesh{structured-grid} \btMethod{eulerian} \btMethod{finite-volume} \btProgLang{c} \btProgLang{fortran} &
&
\href{https://www.spec.org/accel2023/Docs/benchmarks/453.clvrleaf.html}{\scalebox{0.8}{\faIcon{link}}}&
\\
& 455.seismic & \btAppDomain{seismic} \btMethod{finite-difference} \btProgLang{fortran} &
&
\href{https://www.spec.org/accel2023/Docs/benchmarks/455.seismic.html}{\scalebox{0.8}{\faIcon{link}}}&
\\
& 456.spF & \btAppDomain{cfd} \btMethod{solver} \btProgLang{fortran} &
&
\href{https://www.spec.org/accel2023/Docs/benchmarks/456.spF.html}{\scalebox{0.8}{\faIcon{link}}}&
\\
& 457.spC & \btAppDomain{cfd} \btMethod{solver} \btProgLang{c} &
&
\href{https://www.spec.org/accel2023/Docs/benchmarks/457.spC.html}{\scalebox{0.8}{\faIcon{link}}}&
\\
& 459.miniGhost & \btMethod{finite-difference} \btProgLang{c} \btProgLang{fortran} &
&
\href{https://www.spec.org/accel2023/Docs/benchmarks/459.miniGhost.html}{\scalebox{0.8}{\faIcon{link}}}&
\\
& 460.ilbdc & \btAppDomain{cfd} \btMethod{eulerian} \btMethod{finite-difference} \btMethod{lbm} \btProgLang{fortran} &
&
\href{https://www.spec.org/accel2023/Docs/benchmarks/460.ilbdc.html}{\scalebox{0.8}{\faIcon{link}}}&
\\
& 463.swim & \btAppDomain{climate} \btMemAcc{high-memory-bandwidth} \btMethod{finite-difference} \btProgLang{fortran} &
&
\href{https://www.spec.org/accel2023/Docs/benchmarks/463.swim.html}{\scalebox{0.8}{\faIcon{link}}}&
\\
& 470.bt & \btMethod{pde} \btProgLang{c} &
&
\href{https://www.spec.org/accel2023/Docs/benchmarks/470.bt.html}{\scalebox{0.8}{\faIcon{link}}}&
\\
\SetCell[c=2]{l} SPEChpc & & \btScale{large-scale} \btScale{multi-node} \btScale{single-node} \btComm{mpi} \btProgMod{openacc} \btProgMod{openmp} \btProgMod{openmp-target} &
&
\href{https://www.spec.org/hpc2021/}{\scalebox{0.8}{\faIcon{link}}}&
SPEChpc collects its benchmarks into suites of different workload sizes: tiny, small, medium, large. Each size targets different number of tasks and higher memory usage. \\
& LBM D2Q37 & \btAppDomain{cfd} \btProgLang{c} &
\licBsdThree &
\href{https://www.spec.org/hpc2021/docs/benchmarks/505.lbm_t.html}{\scalebox{0.8}{\faIcon{link}}}&
\\
& SOMA & \btAppDomain{physics} \btAppDomain{polymers} \btProgLang{c} &
\licGplThree &
\href{https://www.spec.org/hpc2021/docs/benchmarks/513.soma_t.html}{\scalebox{0.8}{\faIcon{link}}}&
\\
& Tealeaf & \btAppDomain{hep} \btAppDomain{physics} \btProgLang{c} &
\licGplThree &
\href{https://www.spec.org/hpc2021/docs/benchmarks/518.tealeaf_t.html}{\scalebox{0.8}{\faIcon{link}}}&
\\
& Cloverleaf & \btAppDomain{hep} \btAppDomain{physics} \btProgLang{fortran} &
\licGplThree &
\href{https://www.spec.org/hpc2021/docs/benchmarks/519.clvleaf_t.html}{\scalebox{0.8}{\faIcon{link}}}&
\\
& Minisweep & \btAppDomain{nuclear} \btAppDomain{physics} \btProgLang{c} &
\licGplThree &
\href{https://www.spec.org/hpc2021/docs/benchmarks/521.miniswp_t.html}{\scalebox{0.8}{\faIcon{link}}}&
License in linked repository is BSD-2-Clause. \\
& POT3D & \btAppDomain{physics} \btAppDomain{solarphysics} \btProgLang{fortran} &
\licApacheTwo &
\href{https://www.spec.org/hpc2021/docs/benchmarks/528.pot3d_t.html}{\scalebox{0.8}{\faIcon{link}}}&
Custom Apache license \\
& SPH-EXA & \btAppDomain{astrophysics} \btAppDomain{cosmology} \btAppDomain{physics} \btProgLang{c++} &
\licMit &
\href{https://www.spec.org/hpc2021/docs/benchmarks/532.sph_exa_t.html}{\scalebox{0.8}{\faIcon{link}}}&
\\
& HPGMG-FV & \btAppDomain{astrophysics} \btAppDomain{combustion} \btAppDomain{cosmology} \btAppDomain{physics} \btProgLang{c} &
&
\href{https://www.spec.org/hpc2021/docs/benchmarks/534.hpgmgfv_t.html}{\scalebox{0.8}{\faIcon{link}}}&
While the SPEC site lists only copyright, the HPMG repository contains a BSD-2-Clause license. \\
& miniWeather & \btAppDomain{climate} \btProgLang{fortran} &
\licBsdTwo &
\href{https://www.spec.org/hpc2021/docs/benchmarks/535.weather_t.html}{\scalebox{0.8}{\faIcon{link}}}&
\\
\SetCell[c=2]{l} UEABS & & \btComm{mpi} &
\licCcByFour &
\href{https://repository.prace-ri.eu/git/UEABS/ueabs}{\scalebox{0.8}{\faIcon{link}}}&
The Unified European Application Benchmark Suite is a set of 13 application codes maintained by PRACE. The last release was in 2022, and it is probably not maintained anymore. \\
& Alya & \btAppDomain{cfd} \btAppDomain{dft} \btProgLang{fortran} \btProgMod{cuda} \btProgMod{openacc} \btProgMod{openmp} &
\licProp &
\href{https://www.bsc.es/research-development/research-areas/engineering-simulations/alya-high-performance-computational}{\scalebox{0.8}{\faIcon{link}}}&
The Alya System is a Computational Mechanics code capable of solving different physics, each one with its own modelization characteristics, in a coupled way. ALYA is written in Fortran 90/95 and parallelized using MPI and OpenMP. \\
& Code\_Saturne & \btAppDomain{cfd} \btAppDomain{hydrodynamics} \btMath{amgx} \btMath{petsc} \btMethod{finite-volume} \btProgLang{c} \btProgLang{c++} \btProgLang{fortran} \btProgLang{python} \btProgMod{cuda} \btProgMod{openmp} \btProgMod{sycl} &
\licGplTwo &
\href{https://github.com/code-saturne/code_saturne}{\scalebox{0.8}{\faIcon{link}}}&
The code solves the Navier-Stokes equations for incompressible/compressible flows using a predictor-corrector technique and multi-grid algorithms. \\
& CP2K & \btAppDomain{dft} \btAppDomain{molecular-dynamics} \btAppDomain{quantum-chemistry} \btProgLang{fortran} \btProgMod{cuda} \btProgMod{openmp} &
\licGpl &
\href{https://www.cp2k.org/}{\scalebox{0.8}{\faIcon{link}}}&
CP2K is a freely available quantum chemistry and solid-state physics software package for performing atomistic simulations. \\
& GADGET & \btAppDomain{astrophysics} \btProgLang{c} \btProgLang{c++} \btProgMod{openmp} &
\licGplThree &
\href{https://wwwmpa.mpa-garching.mpg.de/gadget4/}{\scalebox{0.8}{\faIcon{link}}}&
GADGET-4 is a freely available code for cosmological N-body/SPH simulations on massively parallel computers with distributed memory. Ref.: \cite{gadget_ref} \\
& GPAW & \btAppDomain{dft} \btMath{blas} \btMath{lapack} \btMath{scalapack} \btMethod{dense-linear-algebra} \btMethod{fft} \btProgLang{python} \btProgMod{c} &
\licGplThree &
\href{https://wiki.fysik.dtu.dk/gpaw/}{\scalebox{0.8}{\faIcon{link}}}&
GPAW is a density-functional theory (DFT) program for ab initio electronic structure calculations using the projector  augmented wave method. \\
& GROMACS & \btAppDomain{molecular-dynamics} \btProgLang{c} \btProgLang{c++} \btProgMod{cuda} \btProgMod{openmp} \btProgMod{sycl} &
\licLgplTwoone &
\href{https://www.gromacs.org/}{\scalebox{0.8}{\faIcon{link}}}&
GROMACS is a versatile package to perform molecular dynamics, i.e. simulate the Newtonian equations of motion for systems with hundreds to millions of particles. Ref.: \cite{pracegromacs_ref} \\
& NAMD & \btAppDomain{molecular-dynamics} \btProgLang{c++} \btProgMod{charm++} \btProgMod{cuda} \btProgMod{hip} \btProgMod{openmp} \btProgMod{sycl} &
\licProp &
\href{http://www.ks.uiuc.edu/Research/namd/}{\scalebox{0.8}{\faIcon{link}}}&
NAMD is a widely used molecular dynamics application designed to simulate bio-molecular systems on a wide variety of compute platforms. Ref.: \cite{pracenamd_ref} \\
& NEMO & \btAppDomain{climate} \btProgLang{c++} \btProgMod{cuda} &
\licProp &
\href{https://www.nemo-ocean.eu/}{\scalebox{0.8}{\faIcon{link}}}&
NEMO is a mathematical modelling framework for prediction in ocean and climate sciences. Free software. Ref.: \cite{nemo_ref} \\
& PFARM & \btAppDomain{astrophysics} \btAppDomain{fusion} \btMath{magma} \btMethod{dense-linear-algebra} \btProgLang{fortran} \btProgMod{cuda} \btProgMod{hip} \btProgMod{openmp} &
\licNone &
\href{https://repository.prace-ri.eu/git/UEABS/ueabs/-/tree/r2.2-dev/pfarm}{\scalebox{0.8}{\faIcon{link}}}&
PFARM uses an R-matrix ab-initio approach to calculate electron-atom and electron-molecule collisions data for a wide range of applications including astrophysics and nuclear fusion.  \\
& QCD & \btAppDomain{hep} \btMath{quda} \btProgLang{c} \btProgLang{c++} \btProgMod{cuda} \btProgMod{openmp} &
GPL-3 &
\href{https://repository.prace-ri.eu/git/UEABS/ueabs/-/blob/master/qcd/README.md}{\scalebox{0.8}{\faIcon{link}}}&
The QCD benchmark is, unlike the other benchmarks in the PRACE application benchmark suite, not a full application but a set of 3 parts which are representative of some of the most compute-intensive parts of QCD calculations. \\
& Quantum ESPRESSO & \btAppDomain{dft} \btAppDomain{material-science} \btAppDomain{physics} \btScale{multi-node} \btComm{mpi} \btMath{elpa} \btProgLang{fortran} \btProgMod{cuf} \btProgMod{openacc} &
GPL-3 &
\href{https://www.quantum-espresso.org/}{\scalebox{0.8}{\faIcon{link}}}&
Quantum Espresso is an integrated suite of Open-Source computer codes for electronic-structure calculations and materials modeling at the nanoscale. \\
& SPECFEM3D & \btAppDomain{seismic} \btMethod{sem} \btProgLang{c} \btProgLang{c++} \btProgLang{python} \btProgMod{cuda} \btProgMod{openmp} &
GPL-3 &
\href{https://github.com/SPECFEM/specfem3d}{\scalebox{0.8}{\faIcon{link}}}&
SPECFEM3D simulates three-dimensional global and regional seismic wave propagation based upon the spectral-element method (SEM). Ref.: \cite{specfem3d_ref} \\
& TensorFlow & \btAppDomain{astrophysics} \btMethod{ai} \btProgLang{c} \btProgLang{c++} \btProgLang{python} \btProgMod{cuda} \btProgMod{openmp} &
\licMit &
\href{https://github.com/maxwelltsai/DeepGalaxy}{\scalebox{0.8}{\faIcon{link}}}&
TensorFlow is a popular open-source library for symbolic math and linear algebra, with particular optimisation for neural-networks-based machine learning workflow. \\
\SetCell[c=2]{l} HPL & & \btAppDomain{synthetic} \btScale{large-scale} \btComm{mpi} \btCompPerf{high-fp} \btCompPerf{vectorization} \btMath{blas} \btMethod{dense-linear-algebra} \btProgLang{c} &
\licBsdFour &
\href{https://www.netlib.org/benchmark/hpl/}{\scalebox{0.8}{\faIcon{link}}}&
High Performance Linpack Ref.: \cite{hpl_ref} \\
\SetCell[c=2]{l} HPCG & & \btScale{large-scale} \btComm{mpi} \btMethod{conjugate-gradient} \btMethod{solver} \btMethod{sparse-linear-algebra} \btProgLang{c++} \btProgMod{openmp} &
\licBsdThree &
\href{https://www.hpcg-benchmark.org}{\scalebox{0.8}{\faIcon{link}}}&
High Performance Conjugate Gradients is a complement to Linpack (HPL) Ref.: \cite{hpcg_ref} \\
\SetCell[c=2]{l} PolyBench & & \btScale{single-node} \btProgLang{c} \btProgLang{fortran} &
&
\href{https://sourceforge.net/projects/polybench/}{\scalebox{0.8}{\faIcon{link}}}&
A benchmark suite of 30 numerical computations with static control flow, extracted from operations in various application domains. Last commit in 2018. \\
\SetCell[c=2]{l} STREAM & & \btScale{single-node} \btMemAcc{regular-memory-access} \btProgLang{c} \btProgLang{fortran} &
\licFree &
\href{https://www.cs.virginia.edu/stream/}{\scalebox{0.8}{\faIcon{link}}}&
Simple synthetic benchmark program that measures sustainable memory bandwidth (in MB/s) and the corresponding computation rate for simple vector kernels. Ref.: \cite{stream_ref} \\
\SetCell[c=2]{l} BabelStream & & \btScale{single-node} \btMemAcc{regular-memory-access} \btProgLang{c++} \btProgMod{cuda} \btProgMod{futhark} \btProgMod{hip} \btProgMod{kokkos} \btProgMod{openacc} \btProgMod{opencl} \btProgMod{openmp} \btProgMod{pstl} \btProgMod{raja} \btProgMod{sycl} \btProgMod{tbb} \btProgMod{thrust} &
\licFree &
\href{https://github.com/UoB-HPC/BabelStream}{\scalebox{0.8}{\faIcon{link}}}&
STREAM in many models for many devices; also available: Julia, Rust, Scala, Java Ref.: \cite{babelstream_ref} \\
\SetCell[c=2]{l} PTRANS & & \btAppDomain{synthetic} \btScale{large-scale} \btCommPerf{network-bisection-bandwidth-bound} \btComm{mpi} \btMath{blas} \btMethod{dense-linear-algebra} \btProgLang{c} &
&
\href{https://www.netlib.org/parkbench/html/matrix-kernels.html}{\scalebox{0.8}{\faIcon{link}}}&
Matrix Transpose \\
\SetCell[c=2]{l} RandomAccess & & \btScale{single-node} \btMemAcc{irregular-memory-access} \btMemAcc{random-memory-access} \btProgLang{c} \btProgMod{openmp} &
&
\href{https://hpcchallenge.org/projectsfiles/hpcc/RandomAccess.html}{\scalebox{0.8}{\faIcon{link}}}&
GUPS (Giga Updates Per Second) Ref.: \cite{randomaccess_ref} \\
\SetCell[c=2]{l} FFT & & \btAppDomain{dft} \btScale{large-scale} \btMemAcc{random-memory-access} \btProgLang{fortran} \btProgMod{cuda} \btProgMod{openmp} &
&
\href{http://www.ffte.jp}{\scalebox{0.8}{\faIcon{link}}}&
1d Discrete Fourier Transforms \\
\SetCell[c=2]{l} b\_eff & & \btScale{large-scale} \btCommPerf{network-bandwidth-bound} \btCommPerf{network-latency-bound} \btComm{mpi} &
\licProp &
\href{https://fs.hlrs.de/projects/par/mpi/b_eff}{\scalebox{0.8}{\faIcon{link}}}&
MPI benchmark for measuring effective accumulated bandwidth in a network; several message sizes, communication patterns and methods used. \\
\SetCell[c=2]{l} LCALS & & \btScale{single-node} \btProgLang{c++} &
&
\href{https://proxyapps.exascaleproject.org/app/lcals/}{\scalebox{0.8}{\faIcon{link}}}&
Livermore Compiler Analysis Loop Suite, a collection of loop kernels based, in part, on historical Livermore Loops benchmarks. Original website currently not reachable. \\
\SetCell[c=2]{l} Graph500 & & \btScale{large-scale} \btComm{mpi} \btMethod{graph-traversal} \btProgLang{c++} \btProgMod{openmp} &
&
\href{https://graph500.org}{\scalebox{0.8}{\faIcon{link}}}&
Linpack for graph problems.  Breadth First Search (BFS). \\
\SetCell[c=2]{l} Rodinia (Julia) & & \btScale{single-node} \btProgLang{julia} \btProgMod{cuda} \btProgMod{opencl} \btProgMod{openmp} &
\licFree &
\href{https://github.com/JuliaParallel/rodinia/}{\scalebox{0.8}{\faIcon{link}}}&
Julia-port of the Rodinia benchmarks, with single and multi-threaded implementations and Julia+CUDA; has the Rodinia license (customized BSD-3-Clause) \\
\SetCell[c=2]{l} Rodinia (DPC++) & & \btProgLang{c++} \btProgMod{dpc++} \btProgMod{sycl} &
\licMit &
\href{https://github.com/artecs-group/rodinia-dpct-dpcpp/}{\scalebox{0.8}{\faIcon{link}}}&
DPC++-translation of Rodinia benchmarks (SYCL tag added for visibility) \\
\SetCell[c=2]{l} Rodinia (SYCL) & & \btProgLang{c++} \btProgMod{sycl} &
\licBsdThree &
\href{https://github.com/zjin-lcf/Rodinia_SYCL}{\scalebox{0.8}{\faIcon{link}}}&
SYCL implementations of Rodinia benchmarks, currently deprecated (integrated into/maintained through HeCBench) \\
\SetCell[c=2]{l} OSU Micro-Benchmarks & & \btAppDomain{synthetic} \btScale{large-scale} \btScale{multi-node} \btScale{single-node} \btCommPerf{network-bandwidth-bound} \btCommPerf{network-bisection-bandwidth-bound} \btCommPerf{network-collectives} \btCommPerf{network-latency-bound} \btCommPerf{network-multi-threaded} \btCommPerf{network-nonblocking-collectives} \btCommPerf{network-onesided} \btCommPerf{network-point-to-point} \btComm{mpi} \btComm{nccl} \btComm{openshmem} \btComm{rccl} \btComm{upc} \btComm{upc++} \btCompPerf{atomics} \btMemAcc{managed-memory} \btMemAcc{regular-memory-access} \btProgLang{c} \btProgLang{java} \btProgLang{python} \btProgMod{cuda} \btProgMod{openacc} \btProgMod{rocm} &
\licBsdThree &
\href{https://mvapich.cse.ohio-state.edu/benchmarks/}{\scalebox{0.8}{\faIcon{link}}}&
A vast collection of network-related micro-benchmarks. \\
\SetCell[c=2]{l} SPEC MPI 2007 & & \btScale{large-scale} \btScale{multi-node} \btScale{single-node} \btComm{mpi} \btProgLang{c} \btProgLang{fortran} &
\licProp &
\href{https://www.spec.org/mpi2007/}{\scalebox{0.8}{\faIcon{link}}}&
MPI-targeted benchmark suite. Last update: 2009 (v2.0). It features the following benchmarks: 104.milc, 107.leslie3d, 113.GemsFDTD, 115.fds4, 121.pop2, 122.tachyon, 125.RAxML, 126.lammps, 127.wrf2, 128.GAPgeofem, 129.tera\_tf, 130.socorro, 132.zeusmp2, 137.lu, 142.dmilc, 143.dleslie, 145.lGemsFDTD, 147.l2wrf2 \\
\SetCell[c=2]{l} GPCNet & & \btScale{large-scale} \btCommPerf{network-bandwidth-bound} \btCommPerf{network-bisection-bandwidth-bound} \btCommPerf{network-collectives} \btCommPerf{network-latency-bound} \btComm{mpi} \btProgLang{c} &
\licApacheTwo &
\href{https://github.com/netbench/GPCNET}{\scalebox{0.8}{\faIcon{link}}}&
Global Performance and Congestion Network Tests. Ref.: \cite{gpcnet_ref} \\
\SetCell[c=2]{l} SPEC OMP 2012 & & \btProgLang{c} \btProgLang{c++} \btProgLang{fortran} \btProgMod{openmp} &
\licProp &
\href{https://www.spec.org/omp2012/}{\scalebox{0.8}{\faIcon{link}}}&
OpenMP-focused benchmark, successor of SPEC OMP 2001 benchmark. Includes the following benchmarks: 350.md, 351.bwaves, 352.nab, 357.bt331, 358.botsalgn, 359.botsspar, 360.ilbdc, 362.fma3d, 363.swim, 367.imagick, 370.mgrid331, 371.applu331, 372.smithwa, 376.kdtree \\
\SetCell[c=2]{l} benchFFT & & \btAppDomain{synthetic} \btScale{single-node} \btMethod{fft} \btProgLang{c} &
\licNone &
\href{https://www.fftw.org/benchfft/}{\scalebox{0.8}{\faIcon{link}}}&
FFTW's FFT benchmark \\
\SetCell[c=2]{l} MLPerf HPC & & \btAppDomain{climate} \btAppDomain{cosmology} \btAppDomain{molecular-dynamics} \btScale{large-scale} \btScale{multi-node} \btMethod{ai} \btProgLang{python} \btProgMod{cuda} \btProgMod{pytorch} \btProgMod{rocm} &
\licApacheTwo &
\href{https://mlcommons.org/benchmarks/training-hpc/}{\scalebox{0.8}{\faIcon{link}}}&
MLPerf Training: HPC collects 4 HPC-related benchmarks: climate segmentation (DeepCAM), cosmology parameter prediction (CosmoFlow), quantum molecular modeling (DimeNet++), protein folding (AlphaFold2). \\
\SetCell[c=2]{l} IO500 & & \btScale{large-scale} \btScale{multi-node} \btComm{mpi} \btCompPerf{i-o} \btProgMod{c} &
\licMit &
\href{https://io500.org/}{\scalebox{0.8}{\faIcon{link}}}&
The IO500 benchmark captures user-experienced I/O performance with a variety of workloads. \\
\SetCell[c=2]{l} Fiber Miniapp & & \btAppDomain{cfd} \btAppDomain{chemistry} \btAppDomain{climate} \btAppDomain{hep} \btAppDomain{molecular-dynamics} \btAppDomain{physics} \btComm{mpi} \btMath{blas} \btMath{lapack} \btProgLang{c} \btProgLang{c++} \btProgLang{fortran} &
\licBsdTwo &
\href{https://fiber-miniapp.github.io/}{\scalebox{0.8}{\faIcon{link}}}&
Suite of miniapps with: CCS QCD, FFVC, NICAM-DC, mVMC, NGS Analyzer, MODYLAS, NTChem, FFB. Different licenses. Last update: 2015. \\
\SetCell[c=2]{l} LULESH & & \btAppDomain{cfd} \btAppDomain{physics} \btComm{mpi} \btMesh{unstructured-grid} \btMethod{hydrodynamics} \btMethod{lagrangian} \btProgLang{c} \btProgLang{c++} \btProgMod{cuda} \btProgMod{openacc} \btProgMod{openmp} \btProgMod{openmp-target} \btProgMod{pstl} &
\licNone &
\href{https://asc.llnl.gov/codes/proxy-apps/lulesh}{\scalebox{0.8}{\faIcon{link}}}&
Suite of proxy apps for 3D Lagrangian hydrodynamics on unstructured mesh. Current version is 2, but no update has happened in quite some time. Ref.: \cite{lulesh_ref} \\
\end{longtblr}

%% file: main.bbl
\begin{thebibliography}{30}
\providecommand{\natexlab}[1]{#1}
\providecommand{\url}[1]{\texttt{#1}}
\providecommand{\urlprefix}{URL }
\expandafter\ifx\csname urlstyle\endcsname\relax
  \providecommand{\doi}[1]{DOI:\discretionary{}{}{}#1}\else
  \providecommand{\doi}{DOI:\discretionary{}{}{}\begingroup
  \urlstyle{rm}\Url}\fi

\bibitem[{Beckingsale et~al.(2019{\natexlab{a}})Beckingsale, Scogland, Burmark,
  Hornung, Jones, Killian, Kunen, Pearce, Robinson and Ryujin}]{RAJA-sc19}
Beckingsale DA, Scogland TR, Burmark J, Hornung R, Jones H, Killian W, Kunen
  AJ, Pearce O, Robinson P and Ryujin BS (2019{\natexlab{a}}) {{RAJA}}:
  {{Portable Performance}} for {{Large-Scale Scientific Applications}}.
\newblock In: \emph{2019 {{IEEE}}/{{ACM International Workshop}} on
  {{Performance}}, {{Portability}} and {{Productivity}} in {{HPC}}
  ({{P3HPC}})}. {Denver, CO, USA}: {IEEE}.
\newblock ISBN 978-1-72816-003-0, pp. 71--81.
\newblock \doi{10.1109/P3HPC49587.2019.00012}.

\bibitem[{Beckingsale et~al.(2019{\natexlab{b}})Beckingsale, Scogland, Burmark,
  Hornung, Jones, Killian, Kunen, Pearce, Robinson and Ryujin}]{rajaperf_ref}
Beckingsale DA, Scogland TR, Burmark J, Hornung R, Jones H, Killian W, Kunen
  AJ, Pearce O, Robinson P and Ryujin BS (2019{\natexlab{b}}) Raja: Portable
  performance for large-scale scientific applications.
\newblock In: \emph{2019 IEEE/ACM International Workshop on Performance,
  Portability and Productivity in HPC (P3HPC)}. IEEE.
\newblock \doi{10.1109/p3hpc49587.2019.00012}.
\newblock \urlprefix\url{http://dx.doi.org/10.1109/P3HPC49587.2019.00012}.

\bibitem[{Berendsen et~al.(1995)Berendsen, van~der Spoel and van
  Drunen}]{pracegromacs_ref}
Berendsen H, van~der Spoel D and van Drunen R (1995) Gromacs: A message-passing
  parallel molecular dynamics implementation.
\newblock \emph{Computer Physics Communications} 91(1–3): 43–56.
\newblock \doi{10.1016/0010-4655(95)00042-e}.
\newblock \urlprefix\url{http://dx.doi.org/10.1016/0010-4655(95)00042-E}.

\bibitem[{Breuer et~al.(2022)Breuer, Lührs, Smolenko and
  Wellmann}]{Breuer2022JUBE}
Breuer T, Lührs S, Smolenko A and Wellmann J (2022) {JUBE} ({V}ersion 2.5.1);
  2.5.1.
\newblock \doi{10.5281/ZENODO.7534373}.
\newblock \urlprefix\url{https://juser.fz-juelich.de/record/917408}.

\bibitem[{Che et~al.(2009)Che, Boyer, Meng, Tarjan, Sheaffer, Lee and
  Skadron}]{rodinia_ref}
Che S, Boyer M, Meng J, Tarjan D, Sheaffer JW, Lee SH and Skadron K (2009)
  Rodinia: A benchmark suite for heterogeneous computing.
\newblock In: \emph{2009 IEEE International Symposium on Workload
  Characterization (IISWC)}. IEEE.
\newblock \doi{10.1109/iiswc.2009.5306797}.
\newblock \urlprefix\url{http://dx.doi.org/10.1109/IISWC.2009.5306797}.

\bibitem[{Chunduri et~al.(2019)Chunduri, Groves, Mendygral, Austin, Balma,
  Kandalla, Kumaran, Lockwood, Parker, Warren, Wichmann and
  Wright}]{gpcnet_ref}
Chunduri S, Groves T, Mendygral P, Austin B, Balma J, Kandalla K, Kumaran K,
  Lockwood G, Parker S, Warren S, Wichmann N and Wright N (2019) Gpcnet:
  designing a benchmark suite for inducing and measuring contention in hpc
  networks.
\newblock In: \emph{Proceedings of the International Conference for High
  Performance Computing, Networking, Storage and Analysis}, SC ’19. ACM.
\newblock \doi{10.1145/3295500.3356215}.
\newblock \urlprefix\url{http://dx.doi.org/10.1145/3295500.3356215}.

\bibitem[{Deakin et~al.(2018)Deakin, Price, Martineau and
  Smith}]{babelstream_ref}
Deakin T, Price J, Martineau M and Smith SM (2018) Evaluating attainable memory
  bandwidth of parallel programming models via babelstream.
\newblock \emph{International Journal of Computational Science and Engineering}
  17(3): 247.
\newblock \doi{10.1504/ijcse.2018.095847}.
\newblock \urlprefix\url{http://dx.doi.org/10.1504/IJCSE.2018.095847}.

\bibitem[{Dongarra et~al.(2015)Dongarra, Heroux and Luszczek}]{hpcg_ref}
Dongarra J, Heroux MA and Luszczek P (2015) High-performance conjugate-gradient
  benchmark: A new metric for ranking high-performance computing systems.
\newblock \emph{The International Journal of High Performance Computing
  Applications} 30(1): 3–10.
\newblock \doi{10.1177/1094342015593158}.
\newblock \urlprefix\url{http://dx.doi.org/10.1177/1094342015593158}.

\bibitem[{Dongarra(1992)}]{hpl_ref}
Dongarra JJ (1992) Performance of various computers using standard linear
  equations software.
\newblock \emph{ACM SIGARCH Computer Architecture News} 20(3): 22–44.
\newblock \doi{10.1145/141868.141871}.
\newblock \urlprefix\url{http://dx.doi.org/10.1145/141868.141871}.

\bibitem[{Herten et~al.(2024)Herten, Achilles, Alvarez, Badwaik, Behle, Bode,
  Breuer, Caviedes-Voullieme, Cherti, Dabah, Sayed, Frings, Gonzalez-Nicolas,
  Gregory, Mood, Hater, Jitsev, John, Meinke, Meyer, Mezentsev, Mirus, Nassyr,
  Penke, Rommer, Sinha, Vieth, Stein, Suarez, Willsch and Zhukov}]{jupiter_ref}
Herten A, Achilles S, Alvarez D, Badwaik J, Behle E, Bode M, Breuer T,
  Caviedes-Voullieme D, Cherti M, Dabah A, Sayed SE, Frings W, Gonzalez-Nicolas
  A, Gregory EB, Mood KH, Hater T, Jitsev J, John CM, Meinke JH, Meyer CI,
  Mezentsev P, Mirus JO, Nassyr S, Penke C, Rommer M, Sinha U, Vieth BvS, Stein
  O, Suarez E, Willsch D and Zhukov I (2024) { Application-Driven Exascale: The
  JUPITER Benchmark Suite }.
\newblock In: \emph{SC24: International Conference for High Performance
  Computing, Networking, Storage and Analysis}. Los Alamitos, CA, USA: IEEE
  Computer Society, pp. 1--45.
\newblock \doi{10.1109/SC41406.2024.00038}.
\newblock
  \urlprefix\url{https://doi.ieeecomputersociety.org/10.1109/SC41406.2024.00038}.

\bibitem[{Herten et~al.(2025)Herten, Pearce and
  Guimaraes}]{benchmarksyaml_link}
Herten A, Pearce O and Guimaraes F (2025) {Benchmarks Survey GitHub}.
\newblock \urlprefix\url{https://github.com/FZJ-JSC/benchmark-survey}.

\bibitem[{Hornung et~al.(2011)Hornung, Keasler and Gokhale}]{lulesh_ref}
Hornung R, Keasler J and Gokhale M (2011) \emph{Hydrodynamics challenge
  problem}.
\newblock \doi{10.2172/1117905}.
\newblock \urlprefix\url{http://dx.doi.org/10.2172/1117905}.

\bibitem[{Ihde et~al.(2022)Ihde, Marten, Eleliemy, Poerwawinata, Silva,
  Tolovski, Ciorba and Rabl}]{10.1007/978-3-030-94437-7_7}
Ihde N, Marten P, Eleliemy A, Poerwawinata G, Silva P, Tolovski I, Ciorba FM
  and Rabl T (2022) A survey of big data, high performance computing,
  and machine learning benchmarks.
\newblock In: Nambiar R and Poess M (eds.) \emph{Performance Evaluation and
  Benchmarking}. Cham: Springer International Publishing.
\newblock ISBN 978-3-030-94437-7, pp. 98--118.

\bibitem[{Jacobsen and Bird(2023)}]{ramble-hpctests2023}
Jacobsen D and Bird B (2023) Ramble: A flexible, extensible, and composable
  experimentation framework.
\newblock In: \emph{{HPC Tests Workshop at the ACM/IEEE International
  Conference on High Performance Computing, Network, Storage, and Analysis
  (SC|23)}}. {Denver, CO, USA}: {ACM}.

\bibitem[{Jin and Vetter(2023)}]{hecbench_ref}
Jin Z and Vetter JS (2023) A benchmark suite for improving performance
  portability of the sycl programming model.
\newblock In: \emph{2023 IEEE International Symposium on Performance Analysis
  of Systems and Software (ISPASS)}. IEEE.
\newblock \doi{10.1109/ispass57527.2023.00041}.
\newblock \urlprefix\url{http://dx.doi.org/10.1109/ISPASS57527.2023.00041}.

\bibitem[{Karakasis et~al.(2020)Karakasis, Manitaras, Rusu,
  Sarmiento-P{\'e}rez, Bignamini, Kraushaar, Jocksch, Omlin, Peretti-Pezzi,
  Augusto, Friesen, He, Gerhardt, Cook, You, Khuvis and
  Tomko}]{10.1007/978-3-030-44728-1_3}
Karakasis V, Manitaras T, Rusu VH, Sarmiento-P{\'e}rez R, Bignamini C,
  Kraushaar M, Jocksch A, Omlin S, Peretti-Pezzi G, Augusto JPSC, Friesen B, He
  Y, Gerhardt L, Cook B, You ZQ, Khuvis S and Tomko K (2020) Enabling
  continuous testing of hpc systems using reframe.
\newblock In: Juckeland G and Chandrasekaran S (eds.) \emph{Tools and
  Techniques for High Performance Computing}. Cham: Springer International
  Publishing.
\newblock ISBN 978-3-030-44728-1, pp. 49--68.

\bibitem[{Komatitsch and Tromp(2002)}]{specfem3d_ref}
Komatitsch D and Tromp J (2002) Spectral-element simulations of global seismic
  wave propagation-i. validation.
\newblock \emph{Geophysical Journal International} 149(2): 390–412.
\newblock \doi{10.1046/j.1365-246x.2002.01653.x}.
\newblock \urlprefix\url{http://dx.doi.org/10.1046/j.1365-246X.2002.01653.x}.

\bibitem[{{LANL Pre-Team}(2019)}]{pavilion}
{LANL Pre-Team} (2019) {{Pavilion Framework}}.
\newblock \url{https://github.com/hpc/pavilion2}.

\bibitem[{Luszczek et~al.(2006)Luszczek, Bailey, Dongarra, Kepner, Lucas,
  Rabenseifner and Takahashi}]{hpcc_ref}
Luszczek PR, Bailey DH, Dongarra JJ, Kepner J, Lucas RF, Rabenseifner R and
  Takahashi D (2006) The hpc challenge (hpcc) benchmark suite.
\newblock In: \emph{Proceedings of the 2006 ACM/IEEE conference on
  Supercomputing}, volume 213. p.~1.

\bibitem[{Madec et~al.(2023)Madec, Bell, Blaker, Bricaud, Bruciaferri,
  Castrillo, Calvert, {Jér\^omeme Chanut}, Clementi, Coward, Epicoco, Éthé,
  Ganderton, Harle, Hutchinson, Iovino, Lea, Lovato, Martin, Martin, Mele,
  Martins, Masson, Mathiot, Mele, Mocavero, M\"{u}ller, Nurser, Paronuzzi,
  Peltier, Person, Rousset, Rynders, Samson, Téchené, Vancoppenolle and
  Wilson}]{nemo_ref}
Madec G, Bell M, Blaker A, Bricaud C, Bruciaferri D, Castrillo M, Calvert D,
  {Jér\^omeme Chanut}, Clementi E, Coward A, Epicoco I, Éthé C, Ganderton J,
  Harle J, Hutchinson K, Iovino D, Lea D, Lovato T, Martin M, Martin N, Mele F,
  Martins D, Masson S, Mathiot P, Mele F, Mocavero S, M\"{u}ller S, Nurser AG,
  Paronuzzi S, Peltier M, Person R, Rousset C, Rynders S, Samson G, Téchené
  S, Vancoppenolle M and Wilson C (2023) Nemo ocean engine reference manual
  \doi{10.5281/ZENODO.8167700}.
\newblock \urlprefix\url{https://zenodo.org/record/8167700}.

\bibitem[{Mattson et~al.(2019)Mattson, Cheng, Coleman, Diamos, Micikevicius,
  Patterson, Tang, Wei, Bailis, Bittorf, Brooks, Chen, Dutta, Gupta, Hazelwood,
  Hock, Huang, Ike, Jia, Kang, Kanter, Kumar, Liao, Ma, Narayanan, Oguntebi,
  Pekhimenko, Pentecost, Reddi, Robie, John, Tabaru, Wu, Xu, Yamazaki, Young
  and Zaharia}]{mattson2019mlperf}
Mattson P, Cheng C, Coleman C, Diamos G, Micikevicius P, Patterson D, Tang H,
  Wei GY, Bailis P, Bittorf V, Brooks D, Chen D, Dutta D, Gupta U, Hazelwood K,
  Hock A, Huang X, Ike A, Jia B, Kang D, Kanter D, Kumar N, Liao J, Ma G,
  Narayanan D, Oguntebi T, Pekhimenko G, Pentecost L, Reddi VJ, Robie T, John
  TS, Tabaru T, Wu CJ, Xu L, Yamazaki M, Young C and Zaharia M (2019) {MLPerf
  Training Benchmark}.

\bibitem[{McCalpin(1995)}]{stream_ref}
McCalpin JD (1995) Memory bandwidth and machine balance in current high
  performance computers.
\newblock \emph{IEEE Computer Society Technical Committee on Computer
  Architecture (TCCA) Newsletter} : 19--25.

\bibitem[{{ML Commons}(2023)}]{mlperf}
{ML Commons} (2023) {MLPerf}.
\newblock \url{https://mlcommons.org/en/}.

\bibitem[{Pearce et~al.(2025)Pearce, Burmark, Hornung, Bogale, Lumsden,
  McKinsey, Yokelson, Boehme, Brink, Taufer and
  Scogland}]{10.1109/rajaperf2024}
Pearce O, Burmark J, Hornung R, Bogale B, Lumsden I, McKinsey M, Yokelson D,
  Boehme D, Brink S, Taufer M and Scogland T (2025) Raja performance suite:
  Performance portability analysis with caliper and thicket.
\newblock In: \emph{Proceedings of the SC '24 Workshops of the International
  Conference on High Performance Computing, Network, Storage, and Analysis},
  SC-W '24. IEEE Press.
\newblock ISBN 9798350355543, p. 1206–1218.
\newblock \doi{10.1109/SCW63240.2024.00162}.
\newblock \urlprefix\url{https://doi.org/10.1109/SCW63240.2024.00162}.

\bibitem[{Pearce et~al.(2023)Pearce, Scott, Becker, Haque, Hanford, Brink,
  Jacobsen, Poxon, Domke and Gamblin}]{hpctests2023-benchpark}
Pearce O, Scott A, Becker G, Haque R, Hanford N, Brink S, Jacobsen D, Poxon H,
  Domke J and Gamblin T (2023) Towards collaborative continuous benchmarking
  for {HPC}.
\newblock In: \emph{Proceedings of the SC '23 Workshops of The International
  Conference on High Performance Computing, Network, Storage, and Analysis},
  SC-W '23. New York, NY, USA: Association for Computing Machinery.
\newblock ISBN 9798400707858, p. 627–635.
\newblock \doi{10.1145/3624062.3624135}.
\newblock \urlprefix\url{https://doi.org/10.1145/3624062.3624135}.

\bibitem[{Phillips et~al.(2020)Phillips, Hardy, Maia, Stone, Ribeiro, Bernardi,
  Buch, Fiorin, Hénin, Jiang, McGreevy, Melo, Radak, Skeel, Singharoy, Wang,
  Roux, Aksimentiev, Luthey-Schulten, Kalé, Schulten, Chipot and
  Tajkhorshid}]{pracenamd_ref}
Phillips JC, Hardy DJ, Maia JDC, Stone JE, Ribeiro JV, Bernardi RC, Buch R,
  Fiorin G, Hénin J, Jiang W, McGreevy R, Melo MCR, Radak BK, Skeel RD,
  Singharoy A, Wang Y, Roux B, Aksimentiev A, Luthey-Schulten Z, Kalé LV,
  Schulten K, Chipot C and Tajkhorshid E (2020) Scalable molecular dynamics on
  cpu and gpu architectures with namd.
\newblock \emph{The Journal of Chemical Physics} 153(4).
\newblock \doi{10.1063/5.0014475}.
\newblock \urlprefix\url{http://dx.doi.org/10.1063/5.0014475}.

\bibitem[{Plimpton et~al.(2006)Plimpton, Brightwell, Vaughan, Underwood and
  Davis}]{randomaccess_ref}
Plimpton S, Brightwell R, Vaughan C, Underwood K and Davis M (2006) A simple
  synchronous distributed-memory algorithm for the hpcc randomaccess benchmark.
\newblock In: \emph{2006 IEEE International Conference on Cluster Computing}.
  IEEE.
\newblock \doi{10.1109/clustr.2006.311859}.
\newblock \urlprefix\url{http://dx.doi.org/10.1109/CLUSTR.2006.311859}.

\bibitem[{Springel et~al.(2021)Springel, Pakmor, Zier and
  Reinecke}]{gadget_ref}
Springel V, Pakmor R, Zier O and Reinecke M (2021) Simulating cosmic structure
  formation with the <scp>gadget</scp>-4 code.
\newblock \emph{Monthly Notices of the Royal Astronomical Society} 506(2):
  2871–2949.
\newblock \doi{10.1093/mnras/stab1855}.
\newblock \urlprefix\url{http://dx.doi.org/10.1093/mnras/stab1855}.

\bibitem[{Thiyagalingam et~al.(2022)Thiyagalingam, Shankar, Fox and
  Hey}]{Thiyagalingam_2022}
Thiyagalingam J, Shankar M, Fox G and Hey T (2022) Scientific machine learning
  benchmarks.
\newblock \emph{Nature Reviews Physics} 4(6): 413–420.
\newblock \doi{10.1038/s42254-022-00441-7}.
\newblock \urlprefix\url{http://dx.doi.org/10.1038/s42254-022-00441-7}.

\bibitem[{Zhang et~al.(2019)Zhang, Zha, Lin, Tu, Li, Liang, Wu and
  Lu}]{10.1007/978-3-030-32813-9_5}
Zhang Q, Zha L, Lin J, Tu D, Li M, Liang F, Wu R and Lu X (2019) A survey on
  deep learning benchmarks: Do we still need new ones?
\newblock In: Zheng C and Zhan J (eds.) \emph{Benchmarking, Measuring, and
  Optimizing}. Cham: Springer International Publishing.
\newblock ISBN 978-3-030-32813-9, pp. 36--49.

\end{thebibliography}
